\documentclass[%
 reprint,
superscriptaddress,
preprintnumbers,
 amsmath,amssymb,
 aps,
 prl,
]{revtex4-2}

\usepackage{graphicx}
\usepackage{dcolumn}
\usepackage{bm}
\usepackage{hyperref}
\usepackage{cleveref}
\usepackage{xcolor}
\usepackage{physics}
\usepackage{multirow}
\usepackage{orcidlink}
\usepackage[normalem]{ulem}

\newcommand{\nstep}{n_{\mathrm{step}}}
\newcommand{\DKL}{D_{\mathrm{KL}}}

\newcommand{\Pf}{\mathcal{P}_{\mathrm{f}}}
\newcommand{\Pre}{\mathcal{P}_{\mathrm{r}}}

\begin{document}


\title{Flow-Based Sampling for Entanglement Entropy and the Machine Learning of Defects}


\author{Andrea Bulgarelli\orcidlink{0009-0002-2917-6125}}\email{andrea.bulgarelli@unito.it}\affiliation{Department of Physics, University of Turin and INFN, Turin unit, Via Pietro Giuria 1, I-10125 Turin, Italy}

\author{Elia Cellini\orcidlink{0000-0002-5664-9752}}\affiliation{Department of Physics, University of Turin and INFN, Turin unit, Via Pietro Giuria 1, I-10125 Turin, Italy}

\author{Karl Jansen\orcidlink{0000-0002-1574-7591}}
\affiliation{Computation-Based Science and Technology Research Center, The Cyprus Institute,  Nicosia, Cyprus}
\affiliation{Deutsches Elektronen-Synchrotron DESY, Zeuthen, Germany}

\author{Stefan K\"{u}hn\orcidlink{0000-0001-7693-350X}}
\affiliation{Deutsches Elektronen-Synchrotron DESY, Zeuthen, Germany}

\author{Alessandro Nada\orcidlink{0000-0002-1766-5186}}
\affiliation{Department of Physics, University of Turin and INFN, Turin unit, Via Pietro Giuria 1, I-10125 Turin, Italy}

\author{Shinichi Nakajima\orcidlink{0000-0003-3970-4569}}
\affiliation{Berlin Institute for the Foundations of Learning and Data (BIFOLD), Berlin, Germany}
\affiliation{Machine Learning Group, Technische Universit\"{a}t Berlin, Berlin, Germany}
\affiliation{RIKEN Center for AIP, Tokyo, Japan}

\author{Kim A.~Nicoli\orcidlink{0000-0001-5933-1822}}
\affiliation{Transdisciplinary Research Area (TRA) Matter, University of Bonn, Germany}
\affiliation{Helmholtz Institute for Radiation and Nuclear Physics (HISKP), Bonn, Germany}

\author{Marco Panero\orcidlink{0000-0001-9477-3749}}\affiliation{Department of Physics, University of Turin and INFN, Turin unit, Via Pietro Giuria 1, I-10125 Turin, Italy}
\affiliation{Department of Physics and Helsinki Institute of Physics, PL 64, FIN-00014 University of Helsinki, Finland}

\date{\today}

\begin{abstract}
We introduce a novel technique to numerically calculate R\'enyi entanglement entropies in lattice quantum field theory using generative models. We describe how flow-based approaches can be combined with the replica trick using a custom neural-network architecture around a lattice defect connecting two replicas. Numerical tests for the $\phi^4$ scalar field theory in two and three dimensions demonstrate that our technique outperforms state-of-the-art Monte Carlo calculations, and exhibit a promising scaling with the defect size.
\end{abstract}

\maketitle


\paragraph{Introduction}

Quantum entanglement is a key property of quantum systems, and has profound implications ranging from high-energy physics~\cite{Klebanov:2007ws} to condensed-matter theory~\cite{Vidal:2002rm}, holding an important role in probing quantum phases of matter and quantum phase transitions. Also, as quantum simulations are emerging as a new tool to study quantum phenomena~\cite{Banuls:2019bmf}, characterizing entanglement in quantum many-body systems is increasingly important. For a system with a factorizable Hilbert space and a density matrix $\rho$, the bipartite entanglement between a subsystem $A$ and its complement $B$ can be quantified by the entanglement entropy
\begin{align}
\label{eq:entanglement_entropy}
    S_A = -\Tr(\rho_A\ln\rho_A), \quad \rho_A = \Tr_B\rho,
\end{align}
namely, the von~Neumann entropy of the reduced density matrix $\rho_A$. While the leading term in the entanglement entropy is proportional to the area of the boundary $\partial A$ between $A$ and $B$~\cite{Eisert:2008ur}, and is ultraviolet-divergent in a quantum field theory, 
the derivative of the entanglement entropy with respect to the linear size $l$ of the subsystem $A$ is finite, and is called the entropic c-function~\cite{Casini:2006es,Nishioka:2006gr}:
\begin{align}
    C = \frac{l^{D-1}}{|\partial A|}\frac{\partial S_A}{\partial l},
\end{align}
where $D$ is the number of spacetime dimensions and $|\partial A|$ the area of $\partial A$. As its name suggests, this quantity provides a measure of the effective number of degrees of freedom of a theory~\cite{Zamolodchikov:1986gt, Casini:2004bw, Casini:2012ei,Casini:2017vbe}.

Computing the entanglement entropy directly from eq.~\eqref{eq:entanglement_entropy} is not possible when $\rho_A$ is not known; tensor networks allow for directly accessing $S_A$, but scaling these methods beyond $1+1$ dimensions is often challenging. However, introducing  the R\'enyi entropies~\cite{Vidal:1998re}
\begin{align}
    S_n = \frac{1}{1-n}\ln\Tr\rho^n_A
    \label{Renyi_entropy_definition}
\end{align}
(and the associated entropic c-functions $C_n$), one can obtain $S_A$ as the $n\to 1$ limit of $S_n$. The advantage of R\'enyi entropies is that they can be computed through the replica trick~\cite{Calabrese:2004eu}, expressing the trace in eq.~\eqref{Renyi_entropy_definition} as the partition function of $n$ copies of the original system, joined together in correspondence of the subsystem $A$ (but not of $B$).
The new geometry, analogous to a Riemann surface, is characterized by the presence of a \textit{defect}, namely the boundary between $A$ and $B$, breaking some of the spacetime symmetries of the original theory~\cite{Bianchi:2015liz}. $S_n$ and $C_n$ can then be expressed in terms of ratios of partition functions in this replica geometry, which can be computed either with tensor networks~\cite{Coser:2013qda}, or with classical~\cite{Buividovich:2008kq, Caraglio:2008pk} or quantum Monte Carlo algorithms~\cite{Hastings:2010zka, DEmidio:2019usm, Zhao:2021njg}. Classical Monte Carlo calculations, however, are often inefficient in the estimate of ratios of partition functions, since the latter cannot be computed directly as primary observables.

In recent years, this problem has been tackled by means of  non-equilibrium Monte Carlo calculations based on Jarzynski's equality~\cite{Jarzynski:1996oqb}, successfully applied in calculations of various physical quantities~\cite{Caselle:2016wsw, Caselle:2018kap, Francesconi:2020fgi}, including R\'enyi entropies~\cite{Alba:2016bcp} and entropic c-functions~\cite{Bulgarelli:2023ofi,Bulgarelli:2024onj}.

In parallel with these developments, during the past decade deep generative models have emerged as a new tool to sample Boltzmann-like distributions~\cite{noe2019boltzmann, Wu:2019elz, cranmer2023advances, Tang2024}. Of particular interest are autoregressive neural networks~\cite{NIPS2016_b1301141} and \textit{normalizing flows} (NFs)~\cite{nfreview} which provide access to the exact probability distributions of statistical systems and to unbiased estimators of partition functions from the learned variational distributions~\cite{PhysRevE.101.023304, Nicoli:2020njz}.

The main question we address in this Letter is whether deep generative models can also be used to efficiently study entanglement, for generic quantum field theories regularized on a lattice. As will be shown below, we give an affirmative answer to this question, and we also present a novel framework in which normalizing flows can be used to study defects in large-scale lattice simulations.

\paragraph{Related Work}
In recent years, significant efforts were devoted to develop efficient algorithms to estimate ratios of partition functions. In Refs.~\cite{Bulgarelli:2023ofi,Bulgarelli:2024onj}, entanglement-related quantities were accurately evaluated through non-equilibrium Markov-chain Monte Carlo (NE-MCMC), leading to a thermodynamic and continuum extrapolation of $C_2$ in the confining $\mathbb{Z}_2$ gauge theory in $2+1$ dimensions. 
At the same time, the idea of using NFs to sample Boltzmann distributions rapidly gained momentum: as reviewed in Ref.~\cite{cranmer2023advances}, initial proof-of-concept studies in quantum chemistry~\cite{noe2019boltzmann} and lattice quantum field theory~\cite{PhysRevD.100.034515,Kanwar:2020xzo,Caselle:2023mvh} were soon followed by works addressing computationally more challenging physical systems, such as lattice quantum chromodynamics in $3+1$ dimensions~\cite{PhysRevD.109.094514}.

Following another line of work, Nicoli et al.~\cite{PhysRevE.101.023304,Nicoli:2020njz,PhysRevD.108.114501} demonstrated that asymptotically unbiased estimators of partition functions could be obtained from a trained generative model. 
While NFs are efficient for low-dimensional systems, they generally exhibit poor scalability with the number of degrees of freedom (d.o.f.). Conversely, NE-MCMC has good scaling properties~\cite{Bonanno:2024udh} but often requires a large number of Monte Carlo updates to be effective. To try and combine the advantages of the two methods, \textit{stochastic normalizing flows} (SNFs) were introduced~\cite{wu2020stochastic,Caselle:2022acb}, and recent studies~\cite{Caselle:2024ent,Bulgarelli:2024brv} have indeed shown their improved scalability and effectiveness. 

In the context of entanglement and replica trick, Bia{\l}as et al.~\cite{bialas2024r} employed autoregressive neural networks 
to estimate R\'enyi entropies, acknowledging the poor scaling of this approach for larger systems. In contrast, our proposed method introduces a novel type of coupling layer for normalizing flows that focuses solely on a region near the defect rather than resampling the entire lattice. 
This straightforward yet effective modification enables our approach to substantially reduce the relevant number of d.o.f. and to overcome the well-known scaling limitations of NFs. For the first time, we demonstrate that NFs can be used to estimate thermodynamic observables in large lattices for $(1+1)$- and $(2+1)$-dimensional scalar field theories.\\

\begin{figure}
    \centering
    \includegraphics[width=.8\linewidth,height=.25\textwidth]{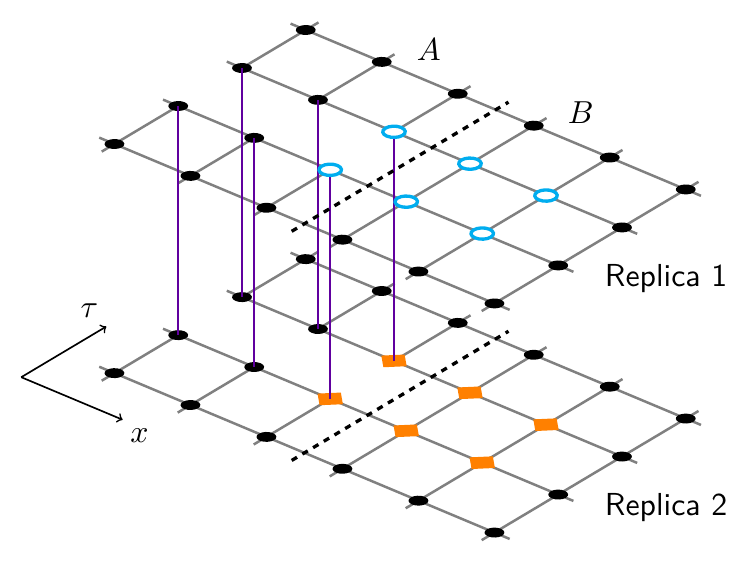}
    \caption{$(1+1)$-dimensional lattice with two replicas ($\tau$ is the Euclidean-time direction). Purple links connect different replicas; dashed lines separate $A$ and $B$. When defect coupling layers act on the configuration, the lattice is divided in three parts: the environment (black sites), which does not enter the coupling layer; frozen sites (empty cyan circles), that are the neural network input; active sites (orange diamonds), which are transformed by the layer.}
    \label{fig:lattice_replica_space}
\end{figure}

\paragraph{Lattice Field Theory and Entanglement Entropy}

We test our proposed methods for a $\phi^4$ real scalar field on a lattice $\Lambda$; the Euclidean action is
\begin{align}
    S = \sum_{x \in \Lambda}  (1-2\lambda)\phi^2(x) + \lambda\phi^4(x)-2\kappa\sum_{\mu=1}^D \phi(x)\phi(x+a\hat{\mu}),
\end{align}
$a$ being the lattice spacing, $\kappa$ the hopping parameter, $\lambda$ the bare quartic coupling, and $\hat{\mu}$ a positively oriented unit vector along the direction $\mu$. The action is invariant under global $\mathbb{Z}_2$ transformations, $\phi\rightarrow \pm\phi$, and the phase diagram of the model is characterized by a critical line of second-order phase transitions in the Ising universality class; therefore, in $1+1$ dimensions, our results can be benchmarked against exact predictions from conformal field theory~\cite{Calabrese:2004eu}.

The calculation of R\'enyi entropies through Monte Carlo simulations is based on the replica trick~\cite{Calabrese:2004eu}: the system is replicated in $n$ independent copies, joined together in the Euclidean-time direction in the subsystem $A$, effectively opening a ``cut'' in the replicated lattice: 
see Fig.~\ref{fig:lattice_replica_space} for a sketch for $n=2$. The trace in eq.~\eqref{Renyi_entropy_definition} is then rewritten as $\Tr\rho_A^n = Z_n/Z^{n}$, where $Z$ is the partition function of the original model and $Z_n$ the partition function of the replicated system. Similarly, the entropic c-function becomes
\begin{align}
\label{eq:cfun}
    C_n = \frac{l^{D-1}}{|\partial A|}\frac{1}{n-1}\lim_{a\rightarrow 0}\frac{1}{a}\ln\frac{Z_n(l)}{Z_n(l+a)}.
\end{align}

\paragraph{Non-Equilibrium MCMC}
Partition-function ratios like the one in eq.~\eqref{eq:cfun} can be computed using Jarzynski's equality~\cite{Jarzynski:1996oqb}: an initial equilibrium distribution $q_0 = \exp(- S_{\mbox{\tiny{i}}})/Z_{\mbox{\tiny{i}}}$ is driven towards a final distribution $p=\exp(- S)/Z$ by a sequence of updates, following a protocol $b(j)$, with a varying transition probability $P_{b(j)}$ that satisfies detailed balance
\begin{equation}
    \phi_0  
    \stackrel{P_{b(1)}}{\longrightarrow} \phi_1  
    \stackrel{P_{b(2)}}{\longrightarrow} \dots
    \stackrel{P_{b(\nstep)}}{\longrightarrow} \phi_{\nstep}
\end{equation}
driving it out of equilibrium. The ratio of partition functions corresponding to the initial and the final distributions is then expressed as
\begin{equation}
\label{eq:jarzynski}
    \frac{Z}{Z_{\mbox{\tiny{i}}}} = \langle \exp(-W) \rangle_{\mathrm{f}},
\end{equation}
where $\langle ... \rangle_{\mathrm{f}}$ denotes an average over a set of such evolutions, and $W$ is the (dimensionless) work done on the system:
\begin{align}
    W &= \sum_{j=0}^{\nstep-1} S_{b(j+1)} (\phi_j) - S_{b(j)} (\phi_j) \\
      &= S(\phi_{\nstep}) - S_{\mbox{\tiny{i}}}(\phi_0) - Q,
\end{align}
with $S_{b(\nstep)} = S$, $S_{b(0)} = S_{\mbox{\tiny{i}}}$, and $Q$ being the (dimensionless) heat. 
The deviation from  equilibrium of a given evolution can be quantified by the Kullback--Leibler (KL) divergence~\cite{kullback1951information} between the forward and reverse transition probabilities:
\begin{equation}
\label{eq:kl_nemcmc}
    \DKL (q_0 \Pf \| p \Pre) = \langle W \rangle_{\mathrm{f}} + \ln \frac{Z}{Z_{\mbox{\tiny{i}}}},
\end{equation}
with $\Pf(\phi_0, \dots, \phi) = \prod_{j=0}^{\nstep-1} P_{j+1} (\phi_j \to \phi_{j+1})$, and $\Pre$ the same for a reversed evolution.
This approach is equivalent to annealed importance sampling~\cite{neal2001annealed}.

\begin{figure*}[t]
    \centering
    \includegraphics[width=\linewidth]{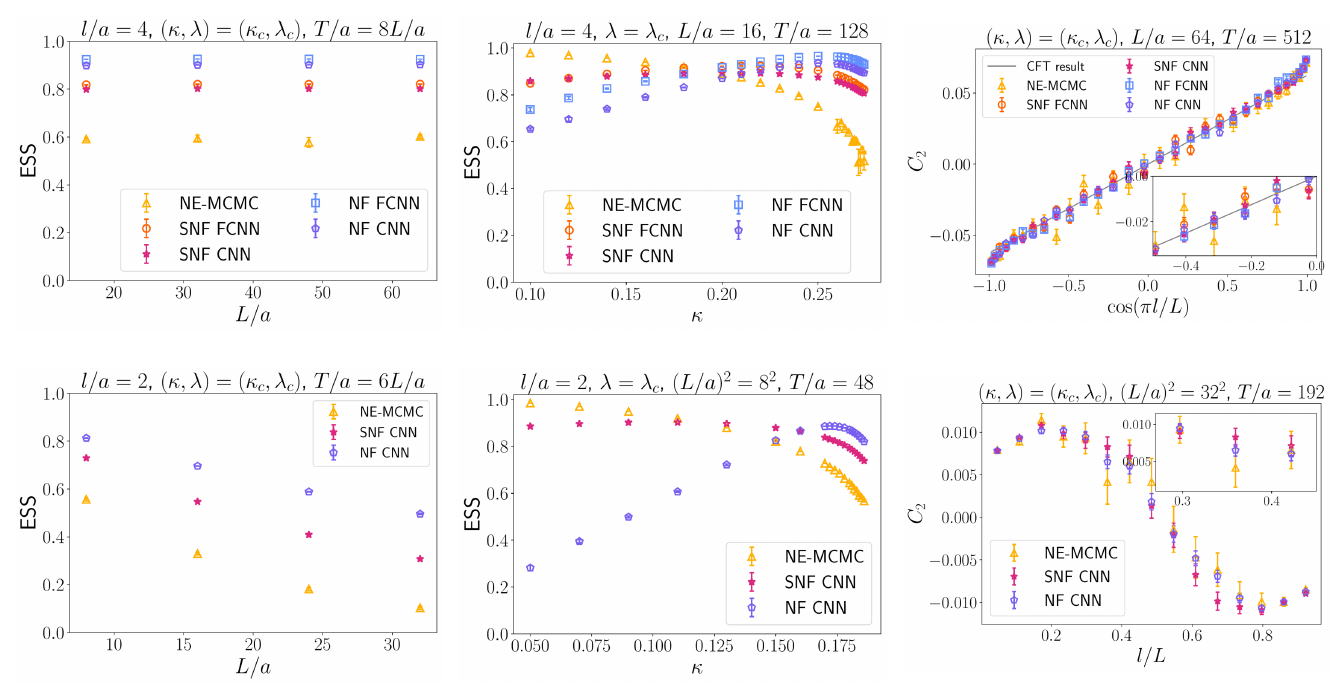}
    \caption{Top row: results in $1+1$ dimensions. Bottom row: results in $2+1$ dimensions. Left panels: quality of the sampling as the model, trained at the smaller value of the volume, is transferred to larger volumes. Central panel: transfer in the hopping parameter $\kappa$ of the model trained at $(\kappa_c, \lambda_c)$. Right panel: estimate of the critical behavior of $C_2$; in $1+1$ dimensions it is compared with the analytical solution~\cite{Calabrese:2004eu}. In all the plots, the quantities in the titles are fixed.}
    \label{fig:results_2d_3d}
\end{figure*}

\paragraph{Normalizing Flows}
Deep generative models can be used to directly estimate partition functions. 
In particular, NFs~\cite{rezende2015variational,nfreview} 
allow one 
to sample highly non-trivial probability distributions. A NF is a composition of  
diffeomorphisms between probability distributions, which can be implemented as coupling layers $g_l$. Concatenating several coupling layers
\begin{equation}
    g_\theta (\phi_0) = \left( g_N \circ \dots \circ g_1 \right) (\phi_0)\,
\end{equation}
one maps a configuration $\phi_0$, sampled from a base distribution $q_0$, into a configuration $\phi=g_\theta(\phi_0)$ 
from the learned distribution $q_\theta$.
Compared to other generative models, NFs 
give access to the exact likelihood of the variational density
\begin{equation}
    q_\theta (\phi) = q_0(g_\theta^{-1}(\phi)) J_{g_\theta}^{-1}
\end{equation}
where $J_{g_\theta}$ is the Jacobian determinant of the transformation $g_\theta$.
By training the parameters of each layer (collectively denoted by $\theta$) 
one can approximate the target distribution $p$ with the variational \textit{Ansatz} $q_\theta$. 
While NFs allow to compute the partition function of a target system~\cite{Nicoli:2020njz}, here we evaluate ratios of partition functions, 
transforming samples from the distribution
$q_0 = \exp(- S_{\mbox{\tiny{i}}})/Z_{\mbox{\tiny{i}}}$ to the target distribution
$p=\exp(- S)/Z$. The variational density $q_\theta$ is constructed by minimizing the KL divergence
\begin{equation}
    \DKL (q_\theta \| p) = \langle -\ln \tilde{w} \rangle_{q_\theta} + \ln \frac{Z}{Z_{\mbox{\tiny{i}}}},
\end{equation}
with respect to $\theta$, and with the weight
\begin{equation}
    \tilde{w} = \exp\left( -(S(\phi) - S_{\mbox{\tiny{i}}}(g_\theta^{-1}(\phi)) - \ln J_{g_\theta}) \right).
\end{equation}
Using $q_0$ as the prior distribution one can show that
\begin{equation}
\label{eq:nf_zz0}
   \frac{Z}{Z_{\mbox{\tiny{i}}}} = \langle \tilde{w} \rangle_{q_\theta}
\end{equation}
where the average $\langle \tilde{w} \rangle_{q_\theta}$ is computed on configurations from the $q_\theta$ distribution. 

\paragraph{Stochastic Normalizing Flows}
The similarity of eqs.~\eqref{eq:jarzynski} and \eqref{eq:nf_zz0} makes it natural to combine the two approaches in a more general architecture, called \textit{stochastic normalizing flows} (SNFs)~\cite{wu2020stochastic,Caselle:2022acb}. In practice, NE-MCMC updates are interleaved with coupling layers,
\begin{equation}
    \phi_0 
    \stackrel{g_1}{\longrightarrow} g_1(\phi_0) 
    \stackrel{P_{b(1)}}{\longrightarrow} \phi_1  
    \stackrel{g_2}{\longrightarrow} \dots
    \stackrel{P_{b(\nstep)}}{\longrightarrow} \phi_{\nstep}.
\end{equation}
Equation~\eqref{eq:jarzynski} is still valid if a generalized work is used
\begin{align}
    W_{\mathrm{SNF}} = S(\phi_{\nstep}) - S_{\mbox{\tiny{i}}}(\phi_0) - Q - \ln J_{g_\theta}
\end{align}
and if the parameters of the coupling layers $g_l$ are trained minimizing the KL divergence of eq.~\eqref{eq:kl_nemcmc} using $W_{\mathrm{SNF}}$.

Crucially, compared to NFs, SNFs benefit from a more favorable scaling with the d.o.f., inherited from the underlying NE-MCMC, as demonstrated in~\cite{Bulgarelli:2024brv}, while it was shown that standard NFs suffer from poor scaling~\cite{deldebbio:2021,Abbott:2023thq,Abbott:2022zsh}. Furthermore, SNFs are generally cheaper than NE-MCMC, since coupling layers are more computationally efficient than Monte Carlo updates.

\paragraph{Proposed Method}
The partition functions in 
eq.~\eqref{eq:cfun} 
are associated with systems that 
differ only for a localized set of links on the lattice. 
This led us to introduce the \textit{defect coupling layer} as a new architecture to study the entropic c-function, where the flow acts on a codimension-two defect, e.g., $(L/a)^{D-2}$. For a pictorial representation see Fig.~\ref{fig:lattice_replica_space}. In the $l$-th coupling layer, while most of the system (the ``environment'') 
is not affected by the transformation, a subset of the lattice is updated by means of a RealNVP affine transformation~\cite{Dinh:2017}
\begin{align}
    \phi_\text{active}^{l+1}= \exp(-|s(\phi_{\text{frozen}}^l)|)\phi_{\text{active}}^l + t(\phi_{\text{frozen}}^l),
    \label{eq:real_NVP}
\end{align}
where $s$ and $t$ are the outputs of an odd neural network, enforcing the $\mathbb{Z}_2$ equivariance~\cite{Nicoli:2020njz,deldebbio:2021}. A related architecture has been discussed in Ref.~\cite{Abbott:2024mix}, with a different active-frozen partitioning;
here, we exploit an even-odd replica decomposition, transforming one replica at a time while keeping the others frozen. The neural networks we used are fully connected (FCNN) and convolutional (CNN).

\paragraph{Numerical Tests}
In what follows, we consider NE-MCMC as the state-of-the-art baseline to benchmark our method. Although SNFs feature better scaling for larger volumes, plain NFs are known to be more computationally efficient for smaller systems; therefore, it is worth investigating both approaches.
We compare the various methods by studying their effective sample size (ESS)
\begin{align}
    \text{ESS} = \frac{\langle e^{-w} \rangle^2}{\langle e^{-2w} \rangle},
\end{align}
with $w=W$ for NE-MCMC, $w=W_{\text{SNF}}$ for SNF and $w=-\ln\tilde{w}$ for NF. Note that $0\leq \text{ESS} \leq 1$, and that the larger the ESS, the smaller the variance of the estimator of $Z/Z_{\mbox{\tiny{i}}}$~\cite{Nicoli:2020njz, Bonanno:2024udh}.

Figure~\ref{fig:results_2d_3d}, top row, displays a comparison 
of architectures that were trained to compute the ratio~\eqref{eq:cfun} with $l/a=1$, for the scalar theory in $1+1$ dimensions on a lattice of sizes $T/a\times L/a = 128\times 16$ and for a critical value of the couplings, $(\kappa_c, \lambda_c) = (0.2758297, 0.03)$~\cite{Bosetti:2015lsa}. The trained models have been then transferred to other values of the parameters without further retraining. 

For all volumes we simulated and for a large set of couplings close to the critical point, both SNFs and NFs significantly outperform the NE-MCMC, with the NFs being the ones with the largest ESS. As the models are transferred to larger volumes, the ESS remains constant, which is not surprising since the patch where the transformation of eq.~\eqref{eq:real_NVP} acts is independent of the volume of the lattice and effectively encodes all relevant information. This allows one to study $C_2$ with high precision for large volumes without further retraining (see Fig.~\ref{fig:results_2d_3d}, top right panel).
When $\kappa$ is varied, the NFs are still the models with the best performances for a significant range of couplings around the critical point.

The results in $2+1$ dimensions share some similarities with those in $1+1$ dimensions. We trained the model for $l/a=1$, $(L/a)^2=8^2$, $T/a = 32$ and at the critical point $(\kappa_c, \lambda_c) = (0.18670475,0.1)$~\cite{Hasenbusch:1999mw}, and again we transferred without retraining. 
The behavior at different values of $\kappa$ is qualitatively similar to the previous case, with the flow-based algorithm outperforming the NE-MCMC for a large range of couplings; moreover, as the lattice volume is increased, NFs always display the largest ESS, see Fig.~\ref{fig:results_2d_3d}, bottom row. In contrast to the $(1+1)$-dimensional case, however, the ESS decreases with the size of the lattice: this behavior is not unexpected as in $2+1$ dimensions the number of d.o.f. on which the model acts grows as $L/a$. Nonetheless, the ESS of the flow-based samplers is still large enough to perform a high-precision study of $C_2$ for large lattices ($T/a\times L^2 / a^2 = 192\times 32^2$ in the bottom right panel of Fig.~\ref{fig:results_2d_3d}).

\begin{figure}[t]
\centering
\includegraphics[width=\linewidth]{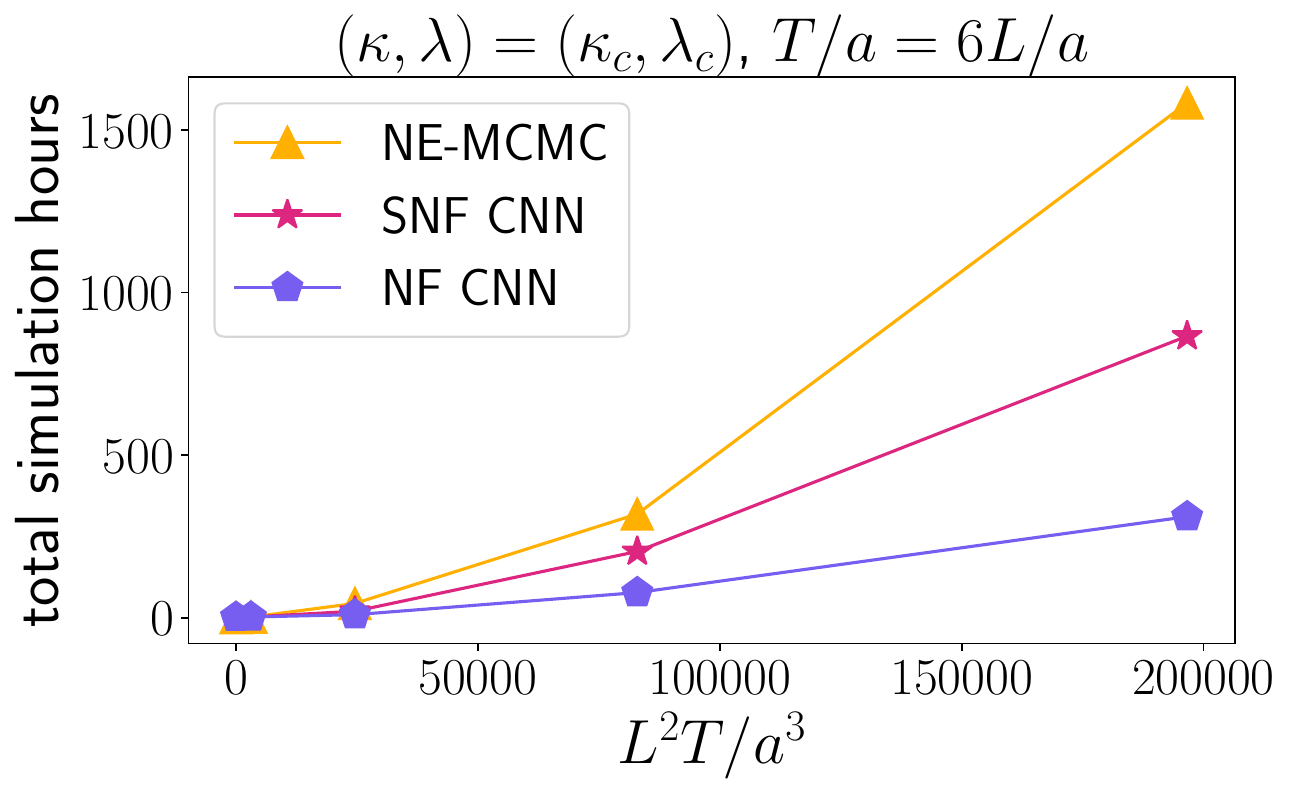}
\caption{Total numerical cost in wall-clock hours to compute the entropic c-function in $2+1$ dimensions for increasingly large volumes at fixed target accuracy. We observe that the larger the volume of the target system, the more evident the advantage. All models used for sampling were trained at the smallest volume, for fixed $l/a=1$ at the critical point.} \label{fig:cost_3d}
\end{figure}

Finally, we studied the total cost for training, thermalization, and sampling, for each method. In Fig.~\ref{fig:cost_3d} we plot the cumulative cost to perform the simulations up to a given lattice volume, to compute the entropic c-function at the critical point in $2+1$ dimensions, similar to the study in Fig.~\ref{fig:results_2d_3d}, bottom right panel. 


This Letter presents, for the first time in the lattice literature, evidence that flow-based methods \textit{outperform} MCMC-based simulations across a physically relevant range of volumes, establishing a new state-of-the-art.
As it is known that SNFs exhibit a more favorable scaling with the degrees of freedom, we expect them to hold promise of being more effective than NFs for larger volumes.
We refer the reader to the supplemental material~\cite{supplemental} for additional results on the scaling for both NFs and SNFs.

\paragraph{Conclusions}

In this Letter, we introduced a novel method to compute physical observables relevant for studying entanglement in quantum systems. Our approach represents a new state-of-the-art method for the numerical evaluation of R\'enyi entropies in the Lagrangian approach. 
Specifically, we proposed a new type of coupling layer for normalizing flows, which acts on a reduced number of degrees of freedom, and is particularly efficient to study lattice defects.

This work is the first example of flow-based sampling outperforming state-of-the-art baselines for large lattices. The scaling of our method in $2+1$ dimensions is especially promising when employing SNFs. The implementation of \textit{direct transfer learning} enables efficient sampling across different system volumes, coupling constant values and defect lengths without the need for retraining. This key advantage allows for a single, inexpensive training on small lattices, facilitating sampling from different configurations of the theory.

Our approach has a broad range of potential applications, including the study of interfaces in spin models and the sampling of different topological sectors in lattice gauge theories. We leave the study of these subjects for future work.
     
\begin{acknowledgments}
    \noindent
    \textit{Acknowledgments} The authors thank M.~Caselle and L.~Funcke for insightful comments and discussions.
    This work is funded by the European Union’s Horizon Europe Framework Programme (HORIZON) under the ERA Chair scheme with grant agreement no.\ 101087126.
    This work is supported with funds from the Ministry of Science, Research and Culture of the State of Brandenburg within the Center for Quantum Technologies and Applications (CQTA).
    A.~N. and E.~C. acknowledge support and A.~B. acknowledges partial support by the Simons Foundation grant 994300 (Simons Collaboration on Confinement and QCD Strings).  A.~N. acknowledges support from the European Union -- Next Generation EU, Mission 4 Component 1, CUP D53D23002970006, under the Italian PRIN ``Progetti di Ricerca di Rilevante Interesse Nazionale – Bando 2022'' prot. 2022ZTPK4E. A.~B., E.~C., A.~N. and M.~P. acknowledge support from the SFT Scientific Initiative of INFN. K.~A.~N. is supported by the Deutsche Forschungsgemeinschaft (DFG, German Research Foundation) as part of the CRC 1639 NuMeriQS – project no. 511713970.
    S.~N. is supported by the German Ministry for Education and Research (BMBF) as BIFOLD – Berlin Institute for the Foundations of Learning and Data (BIFOLD24B),
    and by the European Union’s HORIZON MSCA Doctoral Networks programme project AQTIVATE (101072344).
    The numerical simulations were run on machines of the Consorzio Interuniversitario per il Calcolo Automatico dell'Italia Nord Orientale (CINECA).
\end{acknowledgments}

\nocite{*}

\bibliography{references}

\providecommand{\noopsort}[1]{}\providecommand{\singleletter}[1]{#1}%
\begin{thebibliography}{61}%
\makeatletter
\providecommand \@ifxundefined [1]{%
 \@ifx{#1\undefined}
}%
\providecommand \@ifnum [1]{%
 \ifnum #1\expandafter \@firstoftwo
 \else \expandafter \@secondoftwo
 \fi
}%
\providecommand \@ifx [1]{%
 \ifx #1\expandafter \@firstoftwo
 \else \expandafter \@secondoftwo
 \fi
}%
\providecommand \natexlab [1]{#1}%
\providecommand \enquote  [1]{``#1''}%
\providecommand \bibnamefont  [1]{#1}%
\providecommand \bibfnamefont [1]{#1}%
\providecommand \citenamefont [1]{#1}%
\providecommand \href@noop [0]{\@secondoftwo}%
\providecommand \href [0]{\begingroup \@sanitize@url \@href}%
\providecommand \@href[1]{\@@startlink{#1}\@@href}%
\providecommand \@@href[1]{\endgroup#1\@@endlink}%
\providecommand \@sanitize@url [0]{\catcode `\\12\catcode `\$12\catcode `\&12\catcode `\#12\catcode `\^12\catcode `\_12\catcode `\%12\relax}%
\providecommand \@@startlink[1]{}%
\providecommand \@@endlink[0]{}%
\providecommand \url  [0]{\begingroup\@sanitize@url \@url }%
\providecommand \@url [1]{\endgroup\@href {#1}{\urlprefix }}%
\providecommand \urlprefix  [0]{URL }%
\providecommand \Eprint [0]{\href }%
\providecommand \doibase [0]{http://dx.doi.org/}%
\providecommand \selectlanguage [0]{\@gobble}%
\providecommand \bibinfo  [0]{\@secondoftwo}%
\providecommand \bibfield  [0]{\@secondoftwo}%
\providecommand \translation [1]{[#1]}%
\providecommand \BibitemOpen [0]{}%
\providecommand \bibitemStop [0]{}%
\providecommand \bibitemNoStop [0]{.\EOS\space}%
\providecommand \EOS [0]{\spacefactor3000\relax}%
\providecommand \BibitemShut  [1]{\csname bibitem#1\endcsname}%
\let\auto@bib@innerbib\@empty
\bibitem [{\citenamefont {Klebanov}\ \emph {et~al.}(2008)\citenamefont {Klebanov}, \citenamefont {Kutasov},\ and\ \citenamefont {Murugan}}]{Klebanov:2007ws}%
  \BibitemOpen
  \bibfield  {author} {\bibinfo {author} {\bibfnamefont {I.~R.}\ \bibnamefont {Klebanov}}, \bibinfo {author} {\bibfnamefont {D.}~\bibnamefont {Kutasov}}, \ and\ \bibinfo {author} {\bibfnamefont {A.}~\bibnamefont {Murugan}},\ }\href {\doibase 10.1016/j.nuclphysb.2007.12.017} {\bibfield  {journal} {\bibinfo  {journal} {Nucl. Phys. B}\ }\textbf {\bibinfo {volume} {796}},\ \bibinfo {pages} {274} (\bibinfo {year} {2008})},\ \Eprint {http://arxiv.org/abs/0709.2140} {arXiv:0709.2140 [hep-th]} \BibitemShut {NoStop}%
\bibitem [{\citenamefont {Vidal}\ \emph {et~al.}(2003)\citenamefont {Vidal}, \citenamefont {Latorre}, \citenamefont {Rico},\ and\ \citenamefont {Kitaev}}]{Vidal:2002rm}%
  \BibitemOpen
  \bibfield  {author} {\bibinfo {author} {\bibfnamefont {G.}~\bibnamefont {Vidal}}, \bibinfo {author} {\bibfnamefont {J.~I.}\ \bibnamefont {Latorre}}, \bibinfo {author} {\bibfnamefont {E.}~\bibnamefont {Rico}}, \ and\ \bibinfo {author} {\bibfnamefont {A.}~\bibnamefont {Kitaev}},\ }\href {\doibase 10.1103/PhysRevLett.90.227902} {\bibfield  {journal} {\bibinfo  {journal} {Phys. Rev. Lett.}\ }\textbf {\bibinfo {volume} {90}},\ \bibinfo {pages} {227902} (\bibinfo {year} {2003})},\ \Eprint {http://arxiv.org/abs/quant-ph/0211074} {arXiv:quant-ph/0211074} \BibitemShut {NoStop}%
\bibitem [{\citenamefont {Ba\~nuls}\ \emph {et~al.}(2020)\citenamefont {Ba\~nuls} \emph {et~al.}}]{Banuls:2019bmf}%
  \BibitemOpen
  \bibfield  {author} {\bibinfo {author} {\bibfnamefont {M.~C.}\ \bibnamefont {Ba\~nuls}} \emph {et~al.},\ }\href {\doibase 10.1140/epjd/e2020-100571-8} {\bibfield  {journal} {\bibinfo  {journal} {Eur. Phys. J. D}\ }\textbf {\bibinfo {volume} {74}},\ \bibinfo {pages} {165} (\bibinfo {year} {2020})},\ \Eprint {http://arxiv.org/abs/1911.00003} {arXiv:1911.00003 [quant-ph]} \BibitemShut {NoStop}%
\bibitem [{\citenamefont {Eisert}\ \emph {et~al.}(2010)\citenamefont {Eisert}, \citenamefont {Cramer},\ and\ \citenamefont {Plenio}}]{Eisert:2008ur}%
  \BibitemOpen
  \bibfield  {author} {\bibinfo {author} {\bibfnamefont {J.}~\bibnamefont {Eisert}}, \bibinfo {author} {\bibfnamefont {M.}~\bibnamefont {Cramer}}, \ and\ \bibinfo {author} {\bibfnamefont {M.~B.}\ \bibnamefont {Plenio}},\ }\href {\doibase 10.1103/RevModPhys.82.277} {\bibfield  {journal} {\bibinfo  {journal} {Rev. Mod. Phys.}\ }\textbf {\bibinfo {volume} {82}},\ \bibinfo {pages} {277} (\bibinfo {year} {2010})},\ \Eprint {http://arxiv.org/abs/0808.3773} {arXiv:0808.3773 [quant-ph]} \BibitemShut {NoStop}%
\bibitem [{\citenamefont {Casini}\ and\ \citenamefont {Huerta}(2007)}]{Casini:2006es}%
  \BibitemOpen
  \bibfield  {author} {\bibinfo {author} {\bibfnamefont {H.}~\bibnamefont {Casini}}\ and\ \bibinfo {author} {\bibfnamefont {M.}~\bibnamefont {Huerta}},\ }\href {\doibase 10.1088/1751-8113/40/25/S57} {\bibfield  {journal} {\bibinfo  {journal} {J. Phys. A}\ }\textbf {\bibinfo {volume} {40}},\ \bibinfo {pages} {7031} (\bibinfo {year} {2007})},\ \Eprint {http://arxiv.org/abs/cond-mat/0610375} {arXiv:cond-mat/0610375} \BibitemShut {NoStop}%
\bibitem [{\citenamefont {Nishioka}\ and\ \citenamefont {Takayanagi}(2007)}]{Nishioka:2006gr}%
  \BibitemOpen
  \bibfield  {author} {\bibinfo {author} {\bibfnamefont {T.}~\bibnamefont {Nishioka}}\ and\ \bibinfo {author} {\bibfnamefont {T.}~\bibnamefont {Takayanagi}},\ }\href {\doibase 10.1088/1126-6708/2007/01/090} {\bibfield  {journal} {\bibinfo  {journal} {JHEP}\ }\textbf {\bibinfo {volume} {01}},\ \bibinfo {pages} {090} (\bibinfo {year} {2007})},\ \Eprint {http://arxiv.org/abs/hep-th/0611035} {arXiv:hep-th/0611035} \BibitemShut {NoStop}%
\bibitem [{\citenamefont {Zamolodchikov}(1986)}]{Zamolodchikov:1986gt}%
  \BibitemOpen
  \bibfield  {author} {\bibinfo {author} {\bibfnamefont {A.~B.}\ \bibnamefont {Zamolodchikov}},\ }\href@noop {} {\bibfield  {journal} {\bibinfo  {journal} {JETP Lett.}\ }\textbf {\bibinfo {volume} {43}},\ \bibinfo {pages} {730} (\bibinfo {year} {1986})},\ \bibinfo {note} {[Pisma Zh. Eksp. Teor. Fiz. {\bf 43}, 565 (1986)]}\BibitemShut {NoStop}%
\bibitem [{\citenamefont {Casini}\ and\ \citenamefont {Huerta}(2004)}]{Casini:2004bw}%
  \BibitemOpen
  \bibfield  {author} {\bibinfo {author} {\bibfnamefont {H.}~\bibnamefont {Casini}}\ and\ \bibinfo {author} {\bibfnamefont {M.}~\bibnamefont {Huerta}},\ }\href {\doibase 10.1016/j.physletb.2004.08.072} {\bibfield  {journal} {\bibinfo  {journal} {Phys. Lett. B}\ }\textbf {\bibinfo {volume} {600}},\ \bibinfo {pages} {142} (\bibinfo {year} {2004})},\ \Eprint {http://arxiv.org/abs/hep-th/0405111} {arXiv:hep-th/0405111} \BibitemShut {NoStop}%
\bibitem [{\citenamefont {Casini}\ and\ \citenamefont {Huerta}(2012)}]{Casini:2012ei}%
  \BibitemOpen
  \bibfield  {author} {\bibinfo {author} {\bibfnamefont {H.}~\bibnamefont {Casini}}\ and\ \bibinfo {author} {\bibfnamefont {M.}~\bibnamefont {Huerta}},\ }\href {\doibase 10.1103/PhysRevD.85.125016} {\bibfield  {journal} {\bibinfo  {journal} {Phys. Rev. D}\ }\textbf {\bibinfo {volume} {85}},\ \bibinfo {pages} {125016} (\bibinfo {year} {2012})},\ \Eprint {http://arxiv.org/abs/1202.5650} {arXiv:1202.5650 [hep-th]} \BibitemShut {NoStop}%
\bibitem [{\citenamefont {Casini}\ \emph {et~al.}(2017)\citenamefont {Casini}, \citenamefont {Test\'e},\ and\ \citenamefont {Torroba}}]{Casini:2017vbe}%
  \BibitemOpen
  \bibfield  {author} {\bibinfo {author} {\bibfnamefont {H.}~\bibnamefont {Casini}}, \bibinfo {author} {\bibfnamefont {E.}~\bibnamefont {Test\'e}}, \ and\ \bibinfo {author} {\bibfnamefont {G.}~\bibnamefont {Torroba}},\ }\href {\doibase 10.1103/PhysRevLett.118.261602} {\bibfield  {journal} {\bibinfo  {journal} {Phys. Rev. Lett.}\ }\textbf {\bibinfo {volume} {118}},\ \bibinfo {pages} {261602} (\bibinfo {year} {2017})},\ \Eprint {http://arxiv.org/abs/1704.01870} {arXiv:1704.01870 [hep-th]} \BibitemShut {NoStop}%
\bibitem [{\citenamefont {Vidal}(2000)}]{Vidal:1998re}%
  \BibitemOpen
  \bibfield  {author} {\bibinfo {author} {\bibfnamefont {G.}~\bibnamefont {Vidal}},\ }\href {\doibase 10.1080/09500340008244048} {\bibfield  {journal} {\bibinfo  {journal} {J. Mod. Opt.}\ }\textbf {\bibinfo {volume} {47}},\ \bibinfo {pages} {355} (\bibinfo {year} {2000})},\ \Eprint {http://arxiv.org/abs/quant-ph/9807077} {arXiv:quant-ph/9807077} \BibitemShut {NoStop}%
\bibitem [{\citenamefont {Calabrese}\ and\ \citenamefont {Cardy}(2004)}]{Calabrese:2004eu}%
  \BibitemOpen
  \bibfield  {author} {\bibinfo {author} {\bibfnamefont {P.}~\bibnamefont {Calabrese}}\ and\ \bibinfo {author} {\bibfnamefont {J.~L.}\ \bibnamefont {Cardy}},\ }\href {\doibase 10.1088/1742-5468/2004/06/P06002} {\bibfield  {journal} {\bibinfo  {journal} {J. Stat. Mech.}\ }\textbf {\bibinfo {volume} {0406}},\ \bibinfo {pages} {P06002} (\bibinfo {year} {2004})},\ \Eprint {http://arxiv.org/abs/hep-th/0405152} {arXiv:hep-th/0405152} \BibitemShut {NoStop}%
\bibitem [{\citenamefont {Bianchi}\ \emph {et~al.}(2016)\citenamefont {Bianchi}, \citenamefont {Meineri}, \citenamefont {Myers},\ and\ \citenamefont {Smolkin}}]{Bianchi:2015liz}%
  \BibitemOpen
  \bibfield  {author} {\bibinfo {author} {\bibfnamefont {L.}~\bibnamefont {Bianchi}}, \bibinfo {author} {\bibfnamefont {M.}~\bibnamefont {Meineri}}, \bibinfo {author} {\bibfnamefont {R.~C.}\ \bibnamefont {Myers}}, \ and\ \bibinfo {author} {\bibfnamefont {M.}~\bibnamefont {Smolkin}},\ }\href {\doibase 10.1007/JHEP07(2016)076} {\bibfield  {journal} {\bibinfo  {journal} {JHEP}\ }\textbf {\bibinfo {volume} {07}},\ \bibinfo {pages} {076} (\bibinfo {year} {2016})},\ \Eprint {http://arxiv.org/abs/1511.06713} {arXiv:1511.06713 [hep-th]} \BibitemShut {NoStop}%
\bibitem [{\citenamefont {Coser}\ \emph {et~al.}(2014)\citenamefont {Coser}, \citenamefont {Tagliacozzo},\ and\ \citenamefont {Tonni}}]{Coser:2013qda}%
  \BibitemOpen
  \bibfield  {author} {\bibinfo {author} {\bibfnamefont {A.}~\bibnamefont {Coser}}, \bibinfo {author} {\bibfnamefont {L.}~\bibnamefont {Tagliacozzo}}, \ and\ \bibinfo {author} {\bibfnamefont {E.}~\bibnamefont {Tonni}},\ }\href {\doibase 10.1088/1742-5468/2014/01/P01008} {\bibfield  {journal} {\bibinfo  {journal} {J. Stat. Mech.}\ }\textbf {\bibinfo {volume} {1401}},\ \bibinfo {pages} {P01008} (\bibinfo {year} {2014})},\ \Eprint {http://arxiv.org/abs/1309.2189} {arXiv:1309.2189 [hep-th]} \BibitemShut {NoStop}%
\bibitem [{\citenamefont {Buividovich}\ and\ \citenamefont {Polikarpov}(2008)}]{Buividovich:2008kq}%
  \BibitemOpen
  \bibfield  {author} {\bibinfo {author} {\bibfnamefont {P.~V.}\ \bibnamefont {Buividovich}}\ and\ \bibinfo {author} {\bibfnamefont {M.~I.}\ \bibnamefont {Polikarpov}},\ }\href {\doibase 10.1016/j.nuclphysb.2008.04.024} {\bibfield  {journal} {\bibinfo  {journal} {Nucl. Phys.}\ }\textbf {\bibinfo {volume} {B802}},\ \bibinfo {pages} {458} (\bibinfo {year} {2008})},\ \Eprint {http://arxiv.org/abs/0802.4247} {arXiv:0802.4247 [hep-lat]} \BibitemShut {NoStop}%
\bibitem [{\citenamefont {Caraglio}\ and\ \citenamefont {Gliozzi}(2008)}]{Caraglio:2008pk}%
  \BibitemOpen
  \bibfield  {author} {\bibinfo {author} {\bibfnamefont {M.}~\bibnamefont {Caraglio}}\ and\ \bibinfo {author} {\bibfnamefont {F.}~\bibnamefont {Gliozzi}},\ }\href {\doibase 10.1088/1126-6708/2008/11/076} {\bibfield  {journal} {\bibinfo  {journal} {JHEP}\ }\textbf {\bibinfo {volume} {11}},\ \bibinfo {pages} {076} (\bibinfo {year} {2008})},\ \Eprint {http://arxiv.org/abs/0808.4094} {arXiv:0808.4094 [hep-th]} \BibitemShut {NoStop}%
\bibitem [{\citenamefont {Hastings}\ \emph {et~al.}(2010)\citenamefont {Hastings}, \citenamefont {Gonz{\'a}lez}, \citenamefont {Kallin},\ and\ \citenamefont {Melko}}]{Hastings:2010zka}%
  \BibitemOpen
  \bibfield  {author} {\bibinfo {author} {\bibfnamefont {M.~B.}\ \bibnamefont {Hastings}}, \bibinfo {author} {\bibfnamefont {I.}~\bibnamefont {Gonz{\'a}lez}}, \bibinfo {author} {\bibfnamefont {A.~B.}\ \bibnamefont {Kallin}}, \ and\ \bibinfo {author} {\bibfnamefont {R.~G.}\ \bibnamefont {Melko}},\ }\href {\doibase 10.1103/PhysRevLett.104.157201} {\bibfield  {journal} {\bibinfo  {journal} {Phys. Rev. Lett.}\ }\textbf {\bibinfo {volume} {104}},\ \bibinfo {pages} {157201} (\bibinfo {year} {2010})},\ \Eprint {http://arxiv.org/abs/1001.2335} {arXiv:1001.2335 [cond-mat.str-el]} \BibitemShut {NoStop}%
\bibitem [{\citenamefont {D'Emidio}(2020)}]{DEmidio:2019usm}%
  \BibitemOpen
  \bibfield  {author} {\bibinfo {author} {\bibfnamefont {J.}~\bibnamefont {D'Emidio}},\ }\href {\doibase 10.1103/PhysRevLett.124.110602} {\bibfield  {journal} {\bibinfo  {journal} {Phys. Rev. Lett.}\ }\textbf {\bibinfo {volume} {124}},\ \bibinfo {pages} {110602} (\bibinfo {year} {2020})},\ \Eprint {http://arxiv.org/abs/1904.05918} {arXiv:1904.05918 [cond-mat.str-el]} \BibitemShut {NoStop}%
\bibitem [{\citenamefont {Zhao}\ \emph {et~al.}(2022)\citenamefont {Zhao}, \citenamefont {Chen}, \citenamefont {Wang}, \citenamefont {Yan}, \citenamefont {Cheng},\ and\ \citenamefont {Meng}}]{Zhao:2021njg}%
  \BibitemOpen
  \bibfield  {author} {\bibinfo {author} {\bibfnamefont {J.}~\bibnamefont {Zhao}}, \bibinfo {author} {\bibfnamefont {B.-B.}\ \bibnamefont {Chen}}, \bibinfo {author} {\bibfnamefont {Y.-C.}\ \bibnamefont {Wang}}, \bibinfo {author} {\bibfnamefont {Z.}~\bibnamefont {Yan}}, \bibinfo {author} {\bibfnamefont {M.}~\bibnamefont {Cheng}}, \ and\ \bibinfo {author} {\bibfnamefont {Z.~Y.}\ \bibnamefont {Meng}},\ }\href {\doibase 10.1038/s41535-022-00476-0} {\bibfield  {journal} {\bibinfo  {journal} {Materials}\ }\textbf {\bibinfo {volume} {7}},\ \bibinfo {pages} {69} (\bibinfo {year} {2022})},\ \Eprint {http://arxiv.org/abs/2112.15178} {arXiv:2112.15178 [cond-mat.str-el]} \BibitemShut {NoStop}%
\bibitem [{\citenamefont {Jarzynski}(1997)}]{Jarzynski:1996oqb}%
  \BibitemOpen
  \bibfield  {author} {\bibinfo {author} {\bibfnamefont {C.}~\bibnamefont {Jarzynski}},\ }\href {\doibase 10.1103/PhysRevLett.78.2690} {\bibfield  {journal} {\bibinfo  {journal} {Phys. Rev. Lett.}\ }\textbf {\bibinfo {volume} {78}},\ \bibinfo {pages} {2690} (\bibinfo {year} {1997})},\ \Eprint {http://arxiv.org/abs/cond-mat/9610209} {arXiv:cond-mat/9610209 [cond-mat]} \BibitemShut {NoStop}%
\bibitem [{\citenamefont {Caselle}\ \emph {et~al.}(2016)\citenamefont {Caselle}, \citenamefont {Costagliola}, \citenamefont {Nada}, \citenamefont {Panero},\ and\ \citenamefont {Toniato}}]{Caselle:2016wsw}%
  \BibitemOpen
  \bibfield  {author} {\bibinfo {author} {\bibfnamefont {M.}~\bibnamefont {Caselle}}, \bibinfo {author} {\bibfnamefont {G.}~\bibnamefont {Costagliola}}, \bibinfo {author} {\bibfnamefont {A.}~\bibnamefont {Nada}}, \bibinfo {author} {\bibfnamefont {M.}~\bibnamefont {Panero}}, \ and\ \bibinfo {author} {\bibfnamefont {A.}~\bibnamefont {Toniato}},\ }\href {\doibase 10.1103/PhysRevD.94.034503} {\bibfield  {journal} {\bibinfo  {journal} {Phys. Rev.}\ }\textbf {\bibinfo {volume} {D94}},\ \bibinfo {pages} {034503} (\bibinfo {year} {2016})},\ \Eprint {http://arxiv.org/abs/1604.05544} {arXiv:1604.05544 [hep-lat]} \BibitemShut {NoStop}%
\bibitem [{\citenamefont {Caselle}\ \emph {et~al.}(2018)\citenamefont {Caselle}, \citenamefont {Nada},\ and\ \citenamefont {Panero}}]{Caselle:2018kap}%
  \BibitemOpen
  \bibfield  {author} {\bibinfo {author} {\bibfnamefont {M.}~\bibnamefont {Caselle}}, \bibinfo {author} {\bibfnamefont {A.}~\bibnamefont {Nada}}, \ and\ \bibinfo {author} {\bibfnamefont {M.}~\bibnamefont {Panero}},\ }\href {\doibase 10.1103/PhysRevD.98.054513} {\bibfield  {journal} {\bibinfo  {journal} {Phys. Rev.}\ }\textbf {\bibinfo {volume} {D98}},\ \bibinfo {pages} {054513} (\bibinfo {year} {2018})},\ \Eprint {http://arxiv.org/abs/1801.03110} {arXiv:1801.03110 [hep-lat]} \BibitemShut {NoStop}%
\bibitem [{\citenamefont {Francesconi}\ \emph {et~al.}(2020)\citenamefont {Francesconi}, \citenamefont {Panero},\ and\ \citenamefont {Preti}}]{Francesconi:2020fgi}%
  \BibitemOpen
  \bibfield  {author} {\bibinfo {author} {\bibfnamefont {O.}~\bibnamefont {Francesconi}}, \bibinfo {author} {\bibfnamefont {M.}~\bibnamefont {Panero}}, \ and\ \bibinfo {author} {\bibfnamefont {D.}~\bibnamefont {Preti}},\ }\href {\doibase 10.1007/JHEP07(2020)233} {\bibfield  {journal} {\bibinfo  {journal} {JHEP}\ }\textbf {\bibinfo {volume} {07}},\ \bibinfo {pages} {233} (\bibinfo {year} {2020})},\ \Eprint {http://arxiv.org/abs/2003.13734} {arXiv:2003.13734 [hep-lat]} \BibitemShut {NoStop}%
\bibitem [{\citenamefont {Alba}(2017)}]{Alba:2016bcp}%
  \BibitemOpen
  \bibfield  {author} {\bibinfo {author} {\bibfnamefont {V.}~\bibnamefont {Alba}},\ }\href {\doibase 10.1103/PhysRevE.95.062132} {\bibfield  {journal} {\bibinfo  {journal} {Phys. Rev.}\ }\textbf {\bibinfo {volume} {E95}},\ \bibinfo {pages} {062132} (\bibinfo {year} {2017})},\ \Eprint {http://arxiv.org/abs/1609.02157} {arXiv:1609.02157 [cond-mat.str-el]} \BibitemShut {NoStop}%
\bibitem [{\citenamefont {Bulgarelli}\ and\ \citenamefont {Panero}(2023)}]{Bulgarelli:2023ofi}%
  \BibitemOpen
  \bibfield  {author} {\bibinfo {author} {\bibfnamefont {A.}~\bibnamefont {Bulgarelli}}\ and\ \bibinfo {author} {\bibfnamefont {M.}~\bibnamefont {Panero}},\ }\href {\doibase 10.1007/JHEP06(2023)030} {\bibfield  {journal} {\bibinfo  {journal} {JHEP}\ }\textbf {\bibinfo {volume} {06}},\ \bibinfo {pages} {030} (\bibinfo {year} {2023})},\ \Eprint {http://arxiv.org/abs/2304.03311} {arXiv:2304.03311 [quant-ph]} \BibitemShut {NoStop}%
\bibitem [{\citenamefont {Bulgarelli}\ and\ \citenamefont {Panero}(2024)}]{Bulgarelli:2024onj}%
  \BibitemOpen
  \bibfield  {author} {\bibinfo {author} {\bibfnamefont {A.}~\bibnamefont {Bulgarelli}}\ and\ \bibinfo {author} {\bibfnamefont {M.}~\bibnamefont {Panero}},\ }\href {\doibase 10.1007/JHEP06(2024)041} {\bibfield  {journal} {\bibinfo  {journal} {JHEP}\ }\textbf {\bibinfo {volume} {06}},\ \bibinfo {pages} {041} (\bibinfo {year} {2024})},\ \Eprint {http://arxiv.org/abs/2404.01987} {arXiv:2404.01987 [quant-ph]} \BibitemShut {NoStop}%
\bibitem [{\citenamefont {No{\'e}}\ \emph {et~al.}(2019)\citenamefont {No{\'e}}, \citenamefont {Olsson}, \citenamefont {K{\"o}hler},\ and\ \citenamefont {Wu}}]{noe2019boltzmann}%
  \BibitemOpen
  \bibfield  {author} {\bibinfo {author} {\bibfnamefont {F.}~\bibnamefont {No{\'e}}}, \bibinfo {author} {\bibfnamefont {S.}~\bibnamefont {Olsson}}, \bibinfo {author} {\bibfnamefont {J.}~\bibnamefont {K{\"o}hler}}, \ and\ \bibinfo {author} {\bibfnamefont {H.}~\bibnamefont {Wu}},\ }\href {\doibase 10.1126/science.aaw1147} {\bibfield  {journal} {\bibinfo  {journal} {Science}\ }\textbf {\bibinfo {volume} {365}},\ \bibinfo {pages} {eaaw1147} (\bibinfo {year} {2019})},\ \Eprint {http://arxiv.org/abs/1812.01729} {arXiv:1812.01729 [stat.ML]} \BibitemShut {NoStop}%
\bibitem [{\citenamefont {Wu}\ \emph {et~al.}(2019)\citenamefont {Wu}, \citenamefont {Wang},\ and\ \citenamefont {Zhang}}]{Wu:2019elz}%
  \BibitemOpen
  \bibfield  {author} {\bibinfo {author} {\bibfnamefont {D.}~\bibnamefont {Wu}}, \bibinfo {author} {\bibfnamefont {L.}~\bibnamefont {Wang}}, \ and\ \bibinfo {author} {\bibfnamefont {P.}~\bibnamefont {Zhang}},\ }\href {\doibase 10.1103/PhysRevLett.122.080602} {\bibfield  {journal} {\bibinfo  {journal} {Phys. Rev. Lett.}\ }\textbf {\bibinfo {volume} {122}},\ \bibinfo {pages} {080602} (\bibinfo {year} {2019})}\BibitemShut {NoStop}%
\bibitem [{\citenamefont {Cranmer}\ \emph {et~al.}(2023)\citenamefont {Cranmer}, \citenamefont {Kanwar}, \citenamefont {Racani\`ere}, \citenamefont {Rezende},\ and\ \citenamefont {Shanahan}}]{cranmer2023advances}%
  \BibitemOpen
  \bibfield  {author} {\bibinfo {author} {\bibfnamefont {K.}~\bibnamefont {Cranmer}}, \bibinfo {author} {\bibfnamefont {G.}~\bibnamefont {Kanwar}}, \bibinfo {author} {\bibfnamefont {S.}~\bibnamefont {Racani\`ere}}, \bibinfo {author} {\bibfnamefont {D.~J.}\ \bibnamefont {Rezende}}, \ and\ \bibinfo {author} {\bibfnamefont {P.~E.}\ \bibnamefont {Shanahan}},\ }\href {\doibase 10.1038/s42254-023-00616-w} {\bibfield  {journal} {\bibinfo  {journal} {Nature Rev. Phys.}\ }\textbf {\bibinfo {volume} {5}},\ \bibinfo {pages} {526} (\bibinfo {year} {2023})},\ \Eprint {http://arxiv.org/abs/2309.01156} {arXiv:2309.01156 [hep-lat]} \BibitemShut {NoStop}%
\bibitem [{\citenamefont {Tang}\ \emph {et~al.}(2024)\citenamefont {Tang}, \citenamefont {Liu}, \citenamefont {Zhang},\ and\ \citenamefont {Zhang}}]{Tang2024}%
  \BibitemOpen
  \bibfield  {author} {\bibinfo {author} {\bibfnamefont {Y.}~\bibnamefont {Tang}}, \bibinfo {author} {\bibfnamefont {J.}~\bibnamefont {Liu}}, \bibinfo {author} {\bibfnamefont {J.}~\bibnamefont {Zhang}}, \ and\ \bibinfo {author} {\bibfnamefont {P.}~\bibnamefont {Zhang}},\ }\href {\doibase 10.1038/s41467-024-45172-8} {\bibfield  {journal} {\bibinfo  {journal} {Nature Communications}\ }\textbf {\bibinfo {volume} {15}},\ \bibinfo {pages} {1117} (\bibinfo {year} {2024})},\ \Eprint {http://arxiv.org/abs/2208.08266} {arXiv:2208.08266 [cond-mat.stat-mech]} \BibitemShut {NoStop}%
\bibitem [{\citenamefont {van~den Oord}\ \emph {et~al.}(2016)\citenamefont {van~den Oord}, \citenamefont {Kalchbrenner}, \citenamefont {Espeholt}, \citenamefont {Kavukcuoglu}, \citenamefont {Vinyals},\ and\ \citenamefont {Graves}}]{NIPS2016_b1301141}%
  \BibitemOpen
  \bibfield  {author} {\bibinfo {author} {\bibfnamefont {A.}~\bibnamefont {van~den Oord}}, \bibinfo {author} {\bibfnamefont {N.}~\bibnamefont {Kalchbrenner}}, \bibinfo {author} {\bibfnamefont {L.}~\bibnamefont {Espeholt}}, \bibinfo {author} {\bibfnamefont {K.}~\bibnamefont {Kavukcuoglu}}, \bibinfo {author} {\bibfnamefont {O.}~\bibnamefont {Vinyals}}, \ and\ \bibinfo {author} {\bibfnamefont {A.}~\bibnamefont {Graves}},\ }in\ \href {https://proceedings.neurips.cc/paper/2016/file/b1301141feffabac455e1f90a7de2054-Paper.pdf} {\emph {\bibinfo {booktitle} {Advances in Neural Information Processing Systems}}},\ Vol.~\bibinfo {volume} {29}\ (\bibinfo {year} {2016})\ p.\ \bibinfo {pages} {4797–4805},\ \Eprint {http://arxiv.org/abs/1606.05328} {arXiv:1606.05328 [cs.CV]} \BibitemShut {NoStop}%
\bibitem [{\citenamefont {Papamakarios}\ \emph {et~al.}(2021)\citenamefont {Papamakarios}, \citenamefont {Nalisnick}, \citenamefont {Rezende}, \citenamefont {Mohamed},\ and\ \citenamefont {Lakshminarayanan}}]{nfreview}%
  \BibitemOpen
  \bibfield  {author} {\bibinfo {author} {\bibfnamefont {G.}~\bibnamefont {Papamakarios}}, \bibinfo {author} {\bibfnamefont {E.}~\bibnamefont {Nalisnick}}, \bibinfo {author} {\bibfnamefont {D.~J.}\ \bibnamefont {Rezende}}, \bibinfo {author} {\bibfnamefont {S.}~\bibnamefont {Mohamed}}, \ and\ \bibinfo {author} {\bibfnamefont {B.}~\bibnamefont {Lakshminarayanan}},\ }\href {https://jmlr.org/papers/volume22/19-1028/19-1028.pdf} {\bibfield  {journal} {\bibinfo  {journal} {J. Mach. Learn. Res.}\ }\textbf {\bibinfo {volume} {22}} (\bibinfo {year} {2021})},\ \Eprint {http://arxiv.org/abs/1912.02762} {arXiv:1912.02762 [stat.ML]} \BibitemShut {NoStop}%
\bibitem [{\citenamefont {Nicoli}\ \emph {et~al.}(2020)\citenamefont {Nicoli}, \citenamefont {Nakajima}, \citenamefont {Strodthoff}, \citenamefont {Samek}, \citenamefont {M\"uller},\ and\ \citenamefont {Kessel}}]{PhysRevE.101.023304}%
  \BibitemOpen
  \bibfield  {author} {\bibinfo {author} {\bibfnamefont {K.~A.}\ \bibnamefont {Nicoli}}, \bibinfo {author} {\bibfnamefont {S.}~\bibnamefont {Nakajima}}, \bibinfo {author} {\bibfnamefont {N.}~\bibnamefont {Strodthoff}}, \bibinfo {author} {\bibfnamefont {W.}~\bibnamefont {Samek}}, \bibinfo {author} {\bibfnamefont {K.-R.}\ \bibnamefont {M\"uller}}, \ and\ \bibinfo {author} {\bibfnamefont {P.}~\bibnamefont {Kessel}},\ }\href {\doibase 10.1103/PhysRevE.101.023304} {\bibfield  {journal} {\bibinfo  {journal} {Phys. Rev. E}\ }\textbf {\bibinfo {volume} {101}},\ \bibinfo {pages} {023304} (\bibinfo {year} {2020})},\ \Eprint {http://arxiv.org/abs/1910.13496} {arXiv:1910.13496 [cond-mat]} \BibitemShut {NoStop}%
\bibitem [{\citenamefont {Nicoli}\ \emph {et~al.}(2021)\citenamefont {Nicoli}, \citenamefont {Anders}, \citenamefont {Funcke}, \citenamefont {Hartung}, \citenamefont {Jansen}, \citenamefont {Kessel}, \citenamefont {Nakajima},\ and\ \citenamefont {Stornati}}]{Nicoli:2020njz}%
  \BibitemOpen
  \bibfield  {author} {\bibinfo {author} {\bibfnamefont {K.~A.}\ \bibnamefont {Nicoli}}, \bibinfo {author} {\bibfnamefont {C.~J.}\ \bibnamefont {Anders}}, \bibinfo {author} {\bibfnamefont {L.}~\bibnamefont {Funcke}}, \bibinfo {author} {\bibfnamefont {T.}~\bibnamefont {Hartung}}, \bibinfo {author} {\bibfnamefont {K.}~\bibnamefont {Jansen}}, \bibinfo {author} {\bibfnamefont {P.}~\bibnamefont {Kessel}}, \bibinfo {author} {\bibfnamefont {S.}~\bibnamefont {Nakajima}}, \ and\ \bibinfo {author} {\bibfnamefont {P.}~\bibnamefont {Stornati}},\ }\href {\doibase 10.1103/PhysRevLett.126.032001} {\bibfield  {journal} {\bibinfo  {journal} {Phys. Rev. Lett.}\ }\textbf {\bibinfo {volume} {126}},\ \bibinfo {pages} {032001} (\bibinfo {year} {2021})},\ \Eprint {http://arxiv.org/abs/2007.07115} {arXiv:2007.07115 [hep-lat]} \BibitemShut {NoStop}%
\bibitem [{\citenamefont {Albergo}\ \emph {et~al.}(2019)\citenamefont {Albergo}, \citenamefont {Kanwar},\ and\ \citenamefont {Shanahan}}]{PhysRevD.100.034515}%
  \BibitemOpen
  \bibfield  {author} {\bibinfo {author} {\bibfnamefont {M.~S.}\ \bibnamefont {Albergo}}, \bibinfo {author} {\bibfnamefont {G.}~\bibnamefont {Kanwar}}, \ and\ \bibinfo {author} {\bibfnamefont {P.~E.}\ \bibnamefont {Shanahan}},\ }\href {\doibase 10.1103/PhysRevD.100.034515} {\bibfield  {journal} {\bibinfo  {journal} {Phys. Rev. D}\ }\textbf {\bibinfo {volume} {100}},\ \bibinfo {pages} {034515} (\bibinfo {year} {2019})},\ \Eprint {http://arxiv.org/abs/1904.12072} {arXiv:1904.12072 [hep-lat]} \BibitemShut {NoStop}%
\bibitem [{\citenamefont {Kanwar}\ \emph {et~al.}(2020)\citenamefont {Kanwar}, \citenamefont {Albergo}, \citenamefont {Boyda}, \citenamefont {Cranmer}, \citenamefont {Hackett}, \citenamefont {Racani\`ere}, \citenamefont {Rezende},\ and\ \citenamefont {Shanahan}}]{Kanwar:2020xzo}%
  \BibitemOpen
  \bibfield  {author} {\bibinfo {author} {\bibfnamefont {G.}~\bibnamefont {Kanwar}}, \bibinfo {author} {\bibfnamefont {M.~S.}\ \bibnamefont {Albergo}}, \bibinfo {author} {\bibfnamefont {D.}~\bibnamefont {Boyda}}, \bibinfo {author} {\bibfnamefont {K.}~\bibnamefont {Cranmer}}, \bibinfo {author} {\bibfnamefont {D.~C.}\ \bibnamefont {Hackett}}, \bibinfo {author} {\bibfnamefont {S.}~\bibnamefont {Racani\`ere}}, \bibinfo {author} {\bibfnamefont {D.~J.}\ \bibnamefont {Rezende}}, \ and\ \bibinfo {author} {\bibfnamefont {P.~E.}\ \bibnamefont {Shanahan}},\ }\href {\doibase 10.1103/PhysRevLett.125.121601} {\bibfield  {journal} {\bibinfo  {journal} {Phys. Rev. Lett.}\ }\textbf {\bibinfo {volume} {125}},\ \bibinfo {pages} {121601} (\bibinfo {year} {2020})},\ \Eprint {http://arxiv.org/abs/2003.06413} {arXiv:2003.06413 [hep-lat]} \BibitemShut {NoStop}%
\bibitem [{\citenamefont {Caselle}\ \emph {et~al.}(2024)\citenamefont {Caselle}, \citenamefont {Cellini},\ and\ \citenamefont {Nada}}]{Caselle:2023mvh}%
  \BibitemOpen
  \bibfield  {author} {\bibinfo {author} {\bibfnamefont {M.}~\bibnamefont {Caselle}}, \bibinfo {author} {\bibfnamefont {E.}~\bibnamefont {Cellini}}, \ and\ \bibinfo {author} {\bibfnamefont {A.}~\bibnamefont {Nada}},\ }\href {\doibase 10.1007/JHEP02(2024)048} {\bibfield  {journal} {\bibinfo  {journal} {JHEP}\ }\textbf {\bibinfo {volume} {02}},\ \bibinfo {pages} {048} (\bibinfo {year} {2024})},\ \Eprint {http://arxiv.org/abs/2307.01107} {arXiv:2307.01107 [hep-lat]} \BibitemShut {NoStop}%
\bibitem [{\citenamefont {Abbott}\ \emph {et~al.}(2024{\natexlab{a}})\citenamefont {Abbott}, \citenamefont {Botev}, \citenamefont {Boyda}, \citenamefont {Hackett}, \citenamefont {Kanwar}, \citenamefont {Racani\`ere}, \citenamefont {Rezende}, \citenamefont {Romero-L\'opez}, \citenamefont {Shanahan},\ and\ \citenamefont {Urban}}]{PhysRevD.109.094514}%
  \BibitemOpen
  \bibfield  {author} {\bibinfo {author} {\bibfnamefont {R.}~\bibnamefont {Abbott}}, \bibinfo {author} {\bibfnamefont {A.}~\bibnamefont {Botev}}, \bibinfo {author} {\bibfnamefont {D.}~\bibnamefont {Boyda}}, \bibinfo {author} {\bibfnamefont {D.~C.}\ \bibnamefont {Hackett}}, \bibinfo {author} {\bibfnamefont {G.}~\bibnamefont {Kanwar}}, \bibinfo {author} {\bibfnamefont {S.}~\bibnamefont {Racani\`ere}}, \bibinfo {author} {\bibfnamefont {D.~J.}\ \bibnamefont {Rezende}}, \bibinfo {author} {\bibfnamefont {F.}~\bibnamefont {Romero-L\'opez}}, \bibinfo {author} {\bibfnamefont {P.~E.}\ \bibnamefont {Shanahan}}, \ and\ \bibinfo {author} {\bibfnamefont {J.~M.}\ \bibnamefont {Urban}},\ }\href {\doibase 10.1103/PhysRevD.109.094514} {\bibfield  {journal} {\bibinfo  {journal} {Phys. Rev. D}\ }\textbf {\bibinfo {volume} {109}},\ \bibinfo {pages} {094514} (\bibinfo {year} {2024}{\natexlab{a}})},\ \Eprint {http://arxiv.org/abs/2401.10874} {arXiv:2401.10874 [hep-lat]} \BibitemShut {NoStop}%
\bibitem [{\citenamefont {Nicoli}\ \emph {et~al.}(2023)\citenamefont {Nicoli}, \citenamefont {Anders}, \citenamefont {Hartung}, \citenamefont {Jansen}, \citenamefont {Kessel},\ and\ \citenamefont {Nakajima}}]{PhysRevD.108.114501}%
  \BibitemOpen
  \bibfield  {author} {\bibinfo {author} {\bibfnamefont {K.~A.}\ \bibnamefont {Nicoli}}, \bibinfo {author} {\bibfnamefont {C.~J.}\ \bibnamefont {Anders}}, \bibinfo {author} {\bibfnamefont {T.}~\bibnamefont {Hartung}}, \bibinfo {author} {\bibfnamefont {K.}~\bibnamefont {Jansen}}, \bibinfo {author} {\bibfnamefont {P.}~\bibnamefont {Kessel}}, \ and\ \bibinfo {author} {\bibfnamefont {S.}~\bibnamefont {Nakajima}},\ }\href {\doibase 10.1103/PhysRevD.108.114501} {\bibfield  {journal} {\bibinfo  {journal} {Phys. Rev. D}\ }\textbf {\bibinfo {volume} {108}},\ \bibinfo {pages} {114501} (\bibinfo {year} {2023})},\ \Eprint {http://arxiv.org/abs/2302.14082} {arXiv:2302.14082 [hep-lat]} \BibitemShut {NoStop}%
\bibitem [{\citenamefont {Bonanno}\ \emph {et~al.}(2024)\citenamefont {Bonanno}, \citenamefont {Nada},\ and\ \citenamefont {Vadacchino}}]{Bonanno:2024udh}%
  \BibitemOpen
  \bibfield  {author} {\bibinfo {author} {\bibfnamefont {C.}~\bibnamefont {Bonanno}}, \bibinfo {author} {\bibfnamefont {A.}~\bibnamefont {Nada}}, \ and\ \bibinfo {author} {\bibfnamefont {D.}~\bibnamefont {Vadacchino}},\ }\href {\doibase 10.1007/JHEP04(2024)126} {\bibfield  {journal} {\bibinfo  {journal} {JHEP}\ }\textbf {\bibinfo {volume} {04}},\ \bibinfo {pages} {126} (\bibinfo {year} {2024})},\ \Eprint {http://arxiv.org/abs/2402.06561} {arXiv:2402.06561 [hep-lat]} \BibitemShut {NoStop}%
\bibitem [{\citenamefont {Wu}\ \emph {et~al.}(2020)\citenamefont {Wu}, \citenamefont {K\"{o}hler},\ and\ \citenamefont {Noe}}]{wu2020stochastic}%
  \BibitemOpen
  \bibfield  {author} {\bibinfo {author} {\bibfnamefont {H.}~\bibnamefont {Wu}}, \bibinfo {author} {\bibfnamefont {J.}~\bibnamefont {K\"{o}hler}}, \ and\ \bibinfo {author} {\bibfnamefont {F.}~\bibnamefont {Noe}},\ }in\ \href {https://proceedings.neurips.cc/paper_files/paper/2020/file/41d80bfc327ef980528426fc810a6d7a-Paper.pdf} {\emph {\bibinfo {booktitle} {Advances in Neural Information Processing Systems}}},\ Vol.~\bibinfo {volume} {33}\ (\bibinfo {year} {2020})\ pp.\ \bibinfo {pages} {5933--5944},\ \Eprint {http://arxiv.org/abs/2002.06707} {arXiv:2002.06707 [stat.ML]} \BibitemShut {NoStop}%
\bibitem [{\citenamefont {Caselle}\ \emph {et~al.}(2022)\citenamefont {Caselle}, \citenamefont {Cellini}, \citenamefont {Nada},\ and\ \citenamefont {Panero}}]{Caselle:2022acb}%
  \BibitemOpen
  \bibfield  {author} {\bibinfo {author} {\bibfnamefont {M.}~\bibnamefont {Caselle}}, \bibinfo {author} {\bibfnamefont {E.}~\bibnamefont {Cellini}}, \bibinfo {author} {\bibfnamefont {A.}~\bibnamefont {Nada}}, \ and\ \bibinfo {author} {\bibfnamefont {M.}~\bibnamefont {Panero}},\ }\href {\doibase 10.1007/JHEP07(2022)015} {\bibfield  {journal} {\bibinfo  {journal} {JHEP}\ }\textbf {\bibinfo {volume} {07}},\ \bibinfo {pages} {015} (\bibinfo {year} {2022})},\ \Eprint {http://arxiv.org/abs/2201.08862} {arXiv:2201.08862 [hep-lat]} \BibitemShut {NoStop}%
\bibitem [{\citenamefont {Caselle}\ \emph {et~al.}(2025)\citenamefont {Caselle}, \citenamefont {Cellini},\ and\ \citenamefont {Nada}}]{Caselle:2024ent}%
  \BibitemOpen
  \bibfield  {author} {\bibinfo {author} {\bibfnamefont {M.}~\bibnamefont {Caselle}}, \bibinfo {author} {\bibfnamefont {E.}~\bibnamefont {Cellini}}, \ and\ \bibinfo {author} {\bibfnamefont {A.}~\bibnamefont {Nada}},\ }\href {\doibase 10.1007/JHEP02(2025)090} {\bibfield  {journal} {\bibinfo  {journal} {JHEP}\ }\textbf {\bibinfo {volume} {02}},\ \bibinfo {pages} {090} (\bibinfo {year} {2025})},\ \Eprint {http://arxiv.org/abs/2409.15937} {arXiv:2409.15937 [hep-lat]} \BibitemShut {NoStop}%
\bibitem [{\citenamefont {Bulgarelli}\ \emph {et~al.}(2024)\citenamefont {Bulgarelli}, \citenamefont {Cellini},\ and\ \citenamefont {Nada}}]{Bulgarelli:2024brv}%
  \BibitemOpen
  \bibfield  {author} {\bibinfo {author} {\bibfnamefont {A.}~\bibnamefont {Bulgarelli}}, \bibinfo {author} {\bibfnamefont {E.}~\bibnamefont {Cellini}}, \ and\ \bibinfo {author} {\bibfnamefont {A.}~\bibnamefont {Nada}},\ }\href@noop {} {\  (\bibinfo {year} {2024})},\ \Eprint {http://arxiv.org/abs/2412.00200} {arXiv:2412.00200 [hep-lat]} \BibitemShut {NoStop}%
\bibitem [{\citenamefont {Bia\l{}as}\ \emph {et~al.}(2024)\citenamefont {Bia\l{}as}, \citenamefont {Korcyl}, \citenamefont {Stebel},\ and\ \citenamefont {Zapolski}}]{bialas2024r}%
  \BibitemOpen
  \bibfield  {author} {\bibinfo {author} {\bibfnamefont {P.}~\bibnamefont {Bia\l{}as}}, \bibinfo {author} {\bibfnamefont {P.}~\bibnamefont {Korcyl}}, \bibinfo {author} {\bibfnamefont {T.}~\bibnamefont {Stebel}}, \ and\ \bibinfo {author} {\bibfnamefont {D.}~\bibnamefont {Zapolski}},\ }\href {\doibase 10.1103/PhysRevE.110.044116} {\bibfield  {journal} {\bibinfo  {journal} {Phys. Rev. E}\ }\textbf {\bibinfo {volume} {110}},\ \bibinfo {pages} {044116} (\bibinfo {year} {2024})},\ \Eprint {http://arxiv.org/abs/2406.06193} {arXiv:2406.06193 [cond-mat.stat-mech]} \BibitemShut {NoStop}%
\bibitem [{\citenamefont {Kullback}\ and\ \citenamefont {Leibler}(1951)}]{kullback1951information}%
  \BibitemOpen
  \bibfield  {author} {\bibinfo {author} {\bibfnamefont {S.}~\bibnamefont {Kullback}}\ and\ \bibinfo {author} {\bibfnamefont {R.~A.}\ \bibnamefont {Leibler}},\ }\href@noop {} {\bibfield  {journal} {\bibinfo  {journal} {The annals of mathematical statistics}\ }\textbf {\bibinfo {volume} {22}},\ \bibinfo {pages} {79} (\bibinfo {year} {1951})}\BibitemShut {NoStop}%
\bibitem [{\citenamefont {Neal}(2001)}]{neal2001annealed}%
  \BibitemOpen
  \bibfield  {author} {\bibinfo {author} {\bibfnamefont {R.~M.}\ \bibnamefont {Neal}},\ }\href {https://arxiv.org/abs/physics/9803008} {\bibfield  {journal} {\bibinfo  {journal} {Statistics and computing}\ }\textbf {\bibinfo {volume} {11}},\ \bibinfo {pages} {125} (\bibinfo {year} {2001})},\ \Eprint {http://arxiv.org/abs/physics/9803008} {arXiv:physics/9803008 [physics.comp-ph]} \BibitemShut {NoStop}%
\bibitem [{\citenamefont {Rezende}\ and\ \citenamefont {Mohamed}(2015)}]{rezende2015variational}%
  \BibitemOpen
  \bibfield  {author} {\bibinfo {author} {\bibfnamefont {D.}~\bibnamefont {Rezende}}\ and\ \bibinfo {author} {\bibfnamefont {S.}~\bibnamefont {Mohamed}},\ }in\ \href {https://proceedings.mlr.press/v37/rezende15.html} {\emph {\bibinfo {booktitle} {Proceedings of the 32nd International Conference on Machine Learning}}},\ Vol.~\bibinfo {volume} {37}\ (\bibinfo  {publisher} {PMLR},\ \bibinfo {year} {2015})\ pp.\ \bibinfo {pages} {1530--1538},\ \Eprint {http://arxiv.org/abs/1505.05770} {arXiv:1505.05770 [stat.ML]} \BibitemShut {NoStop}%
\bibitem [{\citenamefont {Del~Debbio}\ \emph {et~al.}(2021)\citenamefont {Del~Debbio}, \citenamefont {Marsh~Rossney},\ and\ \citenamefont {Wilson}}]{deldebbio:2021}%
  \BibitemOpen
  \bibfield  {author} {\bibinfo {author} {\bibfnamefont {L.}~\bibnamefont {Del~Debbio}}, \bibinfo {author} {\bibfnamefont {J.}~\bibnamefont {Marsh~Rossney}}, \ and\ \bibinfo {author} {\bibfnamefont {M.}~\bibnamefont {Wilson}},\ }\href {\doibase 10.1103/PhysRevD.104.094507} {\bibfield  {journal} {\bibinfo  {journal} {Phys. Rev. D}\ }\textbf {\bibinfo {volume} {104}},\ \bibinfo {pages} {094507} (\bibinfo {year} {2021})},\ \Eprint {http://arxiv.org/abs/2105.12481} {arXiv:2105.12481 [hep-lat]} \BibitemShut {NoStop}%
\bibitem [{\citenamefont {Abbott}\ \emph {et~al.}(2023{\natexlab{a}})\citenamefont {Abbott} \emph {et~al.}}]{Abbott:2023thq}%
  \BibitemOpen
  \bibfield  {author} {\bibinfo {author} {\bibfnamefont {R.}~\bibnamefont {Abbott}} \emph {et~al.},\ }\href@noop {} {\  (\bibinfo {year} {2023}{\natexlab{a}})},\ \Eprint {http://arxiv.org/abs/2305.02402} {arXiv:2305.02402 [hep-lat]} \BibitemShut {NoStop}%
\bibitem [{\citenamefont {Abbott}\ \emph {et~al.}(2023{\natexlab{b}})\citenamefont {Abbott} \emph {et~al.}}]{Abbott:2022zsh}%
  \BibitemOpen
  \bibfield  {author} {\bibinfo {author} {\bibfnamefont {R.}~\bibnamefont {Abbott}} \emph {et~al.},\ }\href {\doibase 10.1140/epja/s10050-023-01154-w} {\bibfield  {journal} {\bibinfo  {journal} {Eur. Phys. J. A}\ }\textbf {\bibinfo {volume} {59}},\ \bibinfo {pages} {257} (\bibinfo {year} {2023}{\natexlab{b}})},\ \Eprint {http://arxiv.org/abs/2211.07541} {arXiv:2211.07541 [hep-lat]} \BibitemShut {NoStop}%
\bibitem [{\citenamefont {Dinh}\ \emph {et~al.}(2017)\citenamefont {Dinh}, \citenamefont {Sohl-Dickstein},\ and\ \citenamefont {Bengio}}]{Dinh:2017}%
  \BibitemOpen
  \bibfield  {author} {\bibinfo {author} {\bibfnamefont {L.}~\bibnamefont {Dinh}}, \bibinfo {author} {\bibfnamefont {J.}~\bibnamefont {Sohl-Dickstein}}, \ and\ \bibinfo {author} {\bibfnamefont {S.}~\bibnamefont {Bengio}},\ }in\ \href@noop {} {\emph {\bibinfo {booktitle} {International Conference on Learning Representations}}}\ (\bibinfo {year} {2017})\ \Eprint {http://arxiv.org/abs/1605.08803} {arXiv:1605.08803 [cs.LG]} \BibitemShut {NoStop}%
\bibitem [{\citenamefont {Abbott}\ \emph {et~al.}(2024{\natexlab{b}})\citenamefont {Abbott}, \citenamefont {Boyda}, \citenamefont {Hackett}, \citenamefont {Kanwar}, \citenamefont {Romero-L\'opez}, \citenamefont {Shanahan}, \citenamefont {Urban},\ and\ \citenamefont {Albergo}}]{Abbott:2024mix}%
  \BibitemOpen
  \bibfield  {author} {\bibinfo {author} {\bibfnamefont {R.}~\bibnamefont {Abbott}}, \bibinfo {author} {\bibfnamefont {D.}~\bibnamefont {Boyda}}, \bibinfo {author} {\bibfnamefont {D.~C.}\ \bibnamefont {Hackett}}, \bibinfo {author} {\bibfnamefont {G.}~\bibnamefont {Kanwar}}, \bibinfo {author} {\bibfnamefont {F.}~\bibnamefont {Romero-L\'opez}}, \bibinfo {author} {\bibfnamefont {P.~E.}\ \bibnamefont {Shanahan}}, \bibinfo {author} {\bibfnamefont {J.~M.}\ \bibnamefont {Urban}}, \ and\ \bibinfo {author} {\bibfnamefont {M.~S.}\ \bibnamefont {Albergo}},\ }\href {\doibase 10.22323/1.453.0011} {\bibfield  {journal} {\bibinfo  {journal} {PoS}\ }\textbf {\bibinfo {volume} {LATTICE2023}},\ \bibinfo {pages} {011} (\bibinfo {year} {2024}{\natexlab{b}})},\ \Eprint {http://arxiv.org/abs/2404.11674} {arXiv:2404.11674 [hep-lat]} \BibitemShut {NoStop}%
\bibitem [{\citenamefont {Bosetti}\ \emph {et~al.}(2015)\citenamefont {Bosetti}, \citenamefont {De~Palma},\ and\ \citenamefont {Guagnelli}}]{Bosetti:2015lsa}%
  \BibitemOpen
  \bibfield  {author} {\bibinfo {author} {\bibfnamefont {P.}~\bibnamefont {Bosetti}}, \bibinfo {author} {\bibfnamefont {B.}~\bibnamefont {De~Palma}}, \ and\ \bibinfo {author} {\bibfnamefont {M.}~\bibnamefont {Guagnelli}},\ }\href {\doibase 10.1103/PhysRevD.92.034509} {\bibfield  {journal} {\bibinfo  {journal} {Phys. Rev. D}\ }\textbf {\bibinfo {volume} {92}},\ \bibinfo {pages} {034509} (\bibinfo {year} {2015})},\ \Eprint {http://arxiv.org/abs/1506.08587} {arXiv:1506.08587 [hep-lat]} \BibitemShut {NoStop}%
\bibitem [{\citenamefont {Hasenbusch}(1999)}]{Hasenbusch:1999mw}%
  \BibitemOpen
  \bibfield  {author} {\bibinfo {author} {\bibfnamefont {M.}~\bibnamefont {Hasenbusch}},\ }\href {\doibase 10.1088/0305-4470/32/26/304} {\bibfield  {journal} {\bibinfo  {journal} {J. Phys. A}\ }\textbf {\bibinfo {volume} {32}},\ \bibinfo {pages} {4851} (\bibinfo {year} {1999})},\ \Eprint {http://arxiv.org/abs/hep-lat/9902026} {arXiv:hep-lat/9902026} \BibitemShut {NoStop}%
\bibitem [{sup()}]{supplemental}%
  \BibitemOpen
  \href@noop {} {}\bibinfo {note} {See supplemental material for details on simulations, algorithm performance across theory parameters, and a preliminary scaling analysis. Supplemental material also includes Refs. \cite{Joswig:2022qfe, Matthews:2022sds, Bulgarelli:2024cqc, Kingma:2014vow, Schuh:2025fky}.}\BibitemShut {Stop}%
\bibitem [{\citenamefont {Joswig}\ \emph {et~al.}(2023)\citenamefont {Joswig}, \citenamefont {Kuberski}, \citenamefont {Kuhlmann},\ and\ \citenamefont {Neuendorf}}]{Joswig:2022qfe}%
  \BibitemOpen
  \bibfield  {author} {\bibinfo {author} {\bibfnamefont {F.}~\bibnamefont {Joswig}}, \bibinfo {author} {\bibfnamefont {S.}~\bibnamefont {Kuberski}}, \bibinfo {author} {\bibfnamefont {J.~T.}\ \bibnamefont {Kuhlmann}}, \ and\ \bibinfo {author} {\bibfnamefont {J.}~\bibnamefont {Neuendorf}},\ }\href {\doibase 10.1016/j.cpc.2023.108750} {\bibfield  {journal} {\bibinfo  {journal} {Comput. Phys. Commun.}\ }\textbf {\bibinfo {volume} {288}},\ \bibinfo {pages} {108750} (\bibinfo {year} {2023})},\ \Eprint {http://arxiv.org/abs/2209.14371} {arXiv:2209.14371 [hep-lat]} \BibitemShut {NoStop}%
\bibitem [{\citenamefont {Matthews}\ \emph {et~al.}(2022)\citenamefont {Matthews}, \citenamefont {Arbel}, \citenamefont {Rezende},\ and\ \citenamefont {Doucet}}]{Matthews:2022sds}%
  \BibitemOpen
  \bibfield  {author} {\bibinfo {author} {\bibfnamefont {A.~G. D.~G.}\ \bibnamefont {Matthews}}, \bibinfo {author} {\bibfnamefont {M.}~\bibnamefont {Arbel}}, \bibinfo {author} {\bibfnamefont {D.~J.}\ \bibnamefont {Rezende}}, \ and\ \bibinfo {author} {\bibfnamefont {A.}~\bibnamefont {Doucet}},\ }in\ \href {https://proceedings.mlr.press/v139/arbel21a.html} {\emph {\bibinfo {booktitle} {International Conference on Machine Learning}}}\ (\bibinfo {organization} {PMLR},\ \bibinfo {year} {2022})\ pp.\ \bibinfo {pages} {15196--15219},\ \Eprint {http://arxiv.org/abs/2201.13117} {arXiv:2201.13117 [stat.ML]} \BibitemShut {NoStop}%
\bibitem [{\citenamefont {Bulgarelli}\ \emph {et~al.}(2025)\citenamefont {Bulgarelli}, \citenamefont {Cellini},\ and\ \citenamefont {Nada}}]{Bulgarelli:2024cqc}%
  \BibitemOpen
  \bibfield  {author} {\bibinfo {author} {\bibfnamefont {A.}~\bibnamefont {Bulgarelli}}, \bibinfo {author} {\bibfnamefont {E.}~\bibnamefont {Cellini}}, \ and\ \bibinfo {author} {\bibfnamefont {A.}~\bibnamefont {Nada}},\ }\href {\doibase 10.22323/1.466.0040} {\bibfield  {journal} {\bibinfo  {journal} {PoS}\ }\textbf {\bibinfo {volume} {LATTICE2024}},\ \bibinfo {pages} {040} (\bibinfo {year} {2025})},\ \Eprint {http://arxiv.org/abs/2409.18861} {arXiv:2409.18861 [hep-lat]} \BibitemShut {NoStop}%
\bibitem [{\citenamefont {Kingma}\ and\ \citenamefont {Ba}(2014)}]{Kingma:2014vow}%
  \BibitemOpen
  \bibfield  {author} {\bibinfo {author} {\bibfnamefont {D.~P.}\ \bibnamefont {Kingma}}\ and\ \bibinfo {author} {\bibfnamefont {J.}~\bibnamefont {Ba}}\ }(\bibinfo {year} {2014})\ \Eprint {http://arxiv.org/abs/1412.6980} {arXiv:1412.6980 [cs.LG]} \BibitemShut {NoStop}%
\bibitem [{\citenamefont {Schuh}\ \emph {et~al.}(2025)\citenamefont {Schuh}, \citenamefont {Kreit}, \citenamefont {Berkowitz}, \citenamefont {Funcke}, \citenamefont {Luu}, \citenamefont {Nicoli},\ and\ \citenamefont {Rodekampb}}]{Schuh:2025fky}%
  \BibitemOpen
  \bibfield  {author} {\bibinfo {author} {\bibfnamefont {D.}~\bibnamefont {Schuh}}, \bibinfo {author} {\bibfnamefont {J.}~\bibnamefont {Kreit}}, \bibinfo {author} {\bibfnamefont {E.}~\bibnamefont {Berkowitz}}, \bibinfo {author} {\bibfnamefont {L.}~\bibnamefont {Funcke}}, \bibinfo {author} {\bibfnamefont {T.}~\bibnamefont {Luu}}, \bibinfo {author} {\bibfnamefont {K.~A.}\ \bibnamefont {Nicoli}}, \ and\ \bibinfo {author} {\bibfnamefont {M.}~\bibnamefont {Rodekampb}},\ }\href {\doibase 10.22323/1.466.0069} {\bibfield  {journal} {\bibinfo  {journal} {PoS}\ }\textbf {\bibinfo {volume} {LATTICE2024}},\ \bibinfo {pages} {069} (\bibinfo {year} {2025})},\ \Eprint {http://arxiv.org/abs/2501.07371} {arXiv:2501.07371 [cond-mat.str-el]} \BibitemShut {NoStop}%
\end{thebibliography}%

\newpage\appendix



\section{Details on the simulations}

\begin{figure}[t]
\centerline{\includegraphics[width=.9\linewidth]{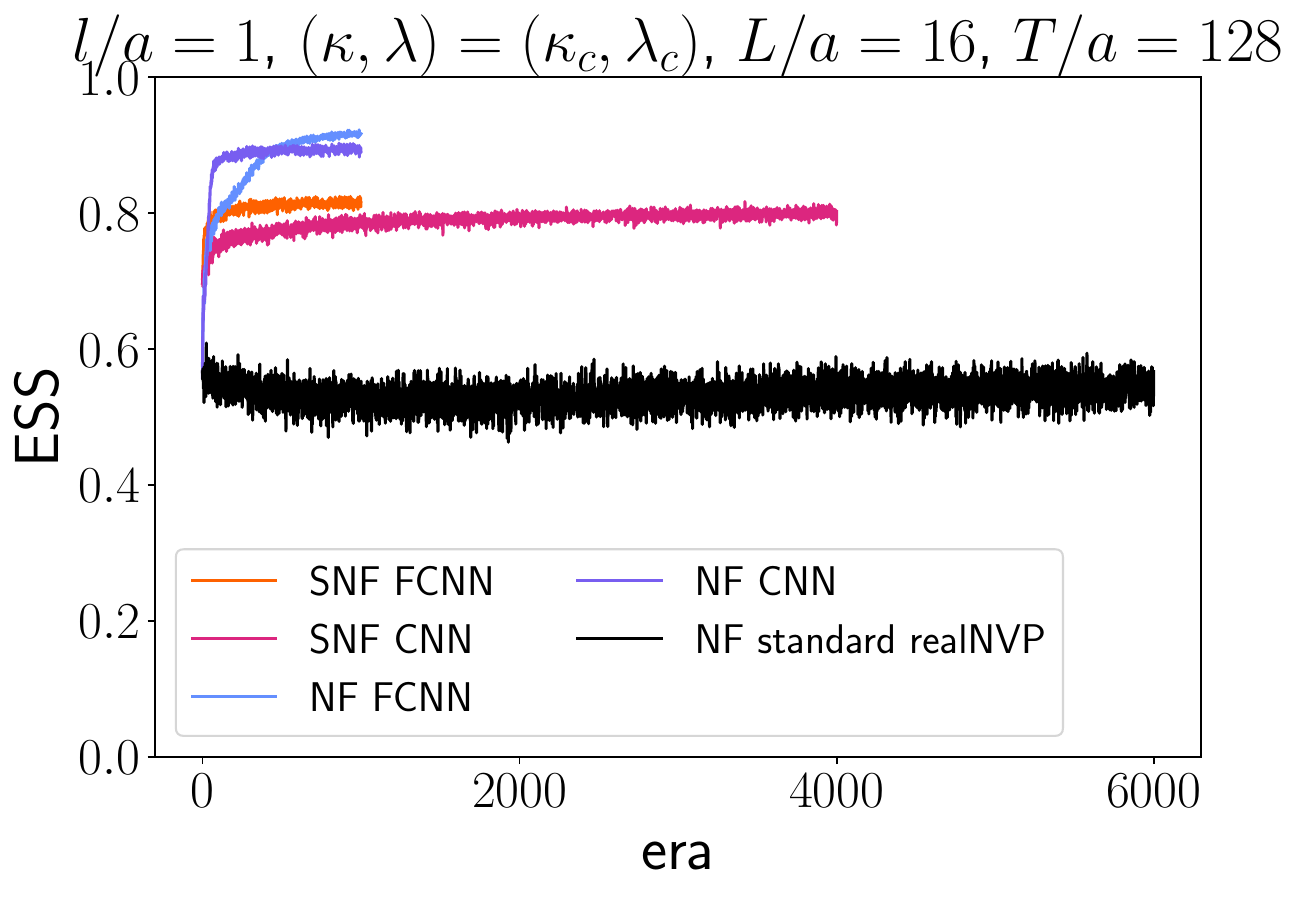}}
\caption{ESS of different models in $D=1+1$ during the training, with a comparison with the standard architecture. In the latter, differently from the defect block, the NF acts on the whole lattice and the coupling layer is made an even-odd masking and a mask on the replicas.} 
\label{fig:comparison_training}
\end{figure}

\textit{Monte Carlo}: The Monte Carlo update used in this work consists of a heatbath, both in the standard MCMC used to sample from the prior and in the stochastic part of the flows. We set the number of steps to thermalize the prior to $10^4$, $5\times 10^4$, $10^5$, $2\times 10^5$ for $T\times L / a^2 = $ $128\times 16$, $256\times 32$, $384\times 48$, $516\times 64$ respectively in $D=1+1$, and $5\times 10^3$, $10^4$, $3\times 10^4$, $5\times 10^4$ steps for $T\times L^2 / a^3=$~$48\times 8^2$, $96\times 16^2$, $144\times 24^2$, $192\times 32^2$ in $D=2+1$. We set to $20$ the number of MCMC steps between two samples of the priors in all cases. To study the performances of the flow-based samplers, we made use of two different conventions. For the plot in Fig.~\ref{fig:results_2d_3d}, top right panel, we fixed the number of samples to $2\times 10^6$, while in the bottom right panel we used $3\times 10^6$ samples, obtaining a different precision for algorithms with different ESS. To study the numerical cost, instead, we fixed the combination $N_{\text{samples}}\left(1/ESS - 1\right)$, which fixes the variance as well as the accuracy; in particular we fixed $N_{\text{samples}} = 2\times 10^6$ for the FCNN-based NF in $D=1+1$ and $N_{\text{samples}}=2\times 10^6$ for the CNN-based NF in $D=2+1$, and in the other cases we vary $N_{\text{samples}}$ accordingly.

\begin{table*}[t]
    \centering
    \begin{tabular}{|c|c|c|c|c|c|c|c|}
    \hline
        & flow & $n_{step}$ & $n_{block}$ (layers per block) & patch size & net (hidden layers, \# kernels) & kernel, stride (CNN only) & activation \\
        \hline\hline
        \multirow{2}{*}{$1+1D$} & SNF & $2$ & $2$ ($2$) & $2\times3$ & FCNN ($0$,/), CNN ($3$, $8$) & $3\times 3$, $1$ , $1$ & Tanh  \\
        \cline{2-8}
          & NF & $0$ & $4$ ($2$) & $2\times3$, $4\times5$ & FCNN ($0$,/), CNN ($3$, $8$) & $3\times 3$, $1$ , $1$ & Tanh  \\
        \hline
        \multirow{2}{*}{$2+1D$} & SNF & $5$ & $5$ ($2$) & $2\times3\times L/a$ & CNN ($3$, $8$) & $3\times 3\times 3$, $1$ , $1$ & Tanh \\
        \cline{2-8}
          & NF & $0$ & $5$ ($2$) & $2\times3\times L/a$ & CNN ($3$, $8$) & $3\times 3\times 3$, $1$ , $1$ & Tanh \\
    \hline
    \end{tabular}
    \caption{Design of the flows analyzed in the main text. When padding is used in CNN, this is always a zero-padding. }
    \label{tab:flows_design}
\end{table*}

\begin{table*}[t]
    \centering
    \begin{tabular}{|c|c|c|c|c|c|c|c|}
    \hline
        & flow & net & training steps & batch size & $l/a$ & ($T/a$, $L/a$) & $(\kappa, \lambda)$ \\
        \hline\hline
        \multirow{4}{*}{$1+1D$} & \multirow{2}{*}{SNF} & FCNN & $10^5$ & $100$ & $1$ & ($128$, $16$) & $(0.2758297, 0.03)$ \\
        \cline{3-8}
          &  & CNN & $4\times 10^5$ & $100$ & $1$ & ($128$, $16$) & $(0.2758297, 0.03)$ \\
        \cline{2-8}
          & \multirow{2}{*}{NF} & FCNN & $10^5$ & $100$ & $1$ & ($128$, $16$) & $(0.2758297, 0.03)$ \\
        \cline{3-8}
        &  & CNN & $10^5$ & $100$ & $1$ & ($128$, $16$) & $(0.2758297, 0.03)$ \\
        \hline
        \multirow{2}{*}{$2+1D$} & SNF & CNN & $2\times 10^5$ & $100$ & $1$ & ($32$, $8$) & $(0.18670475, 0.1)$ \\
        \cline{2-8}
          & NF & CNN & $2\times 10^5$ & $100$ & $1$ & ($32$, $8$) & $(0.18670475, 0.1)$ \\
    \hline
    \end{tabular}
    \caption{Training of the flows analyzed in the main text.}
    \label{tab:flows_training}
\end{table*}

\begin{figure*}[!]
\centerline{\includegraphics[height=0.25\textheight]{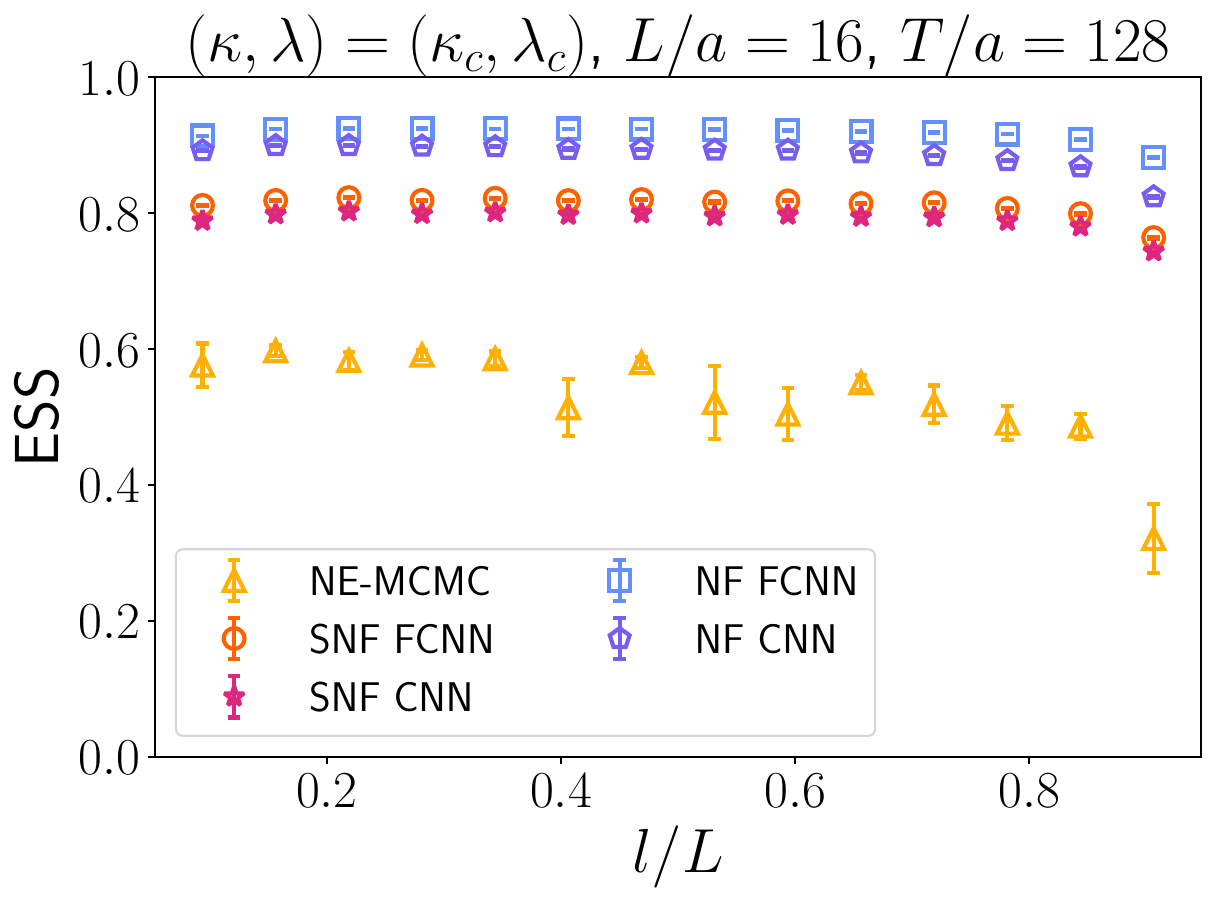}  \includegraphics[height=0.25\textheight]{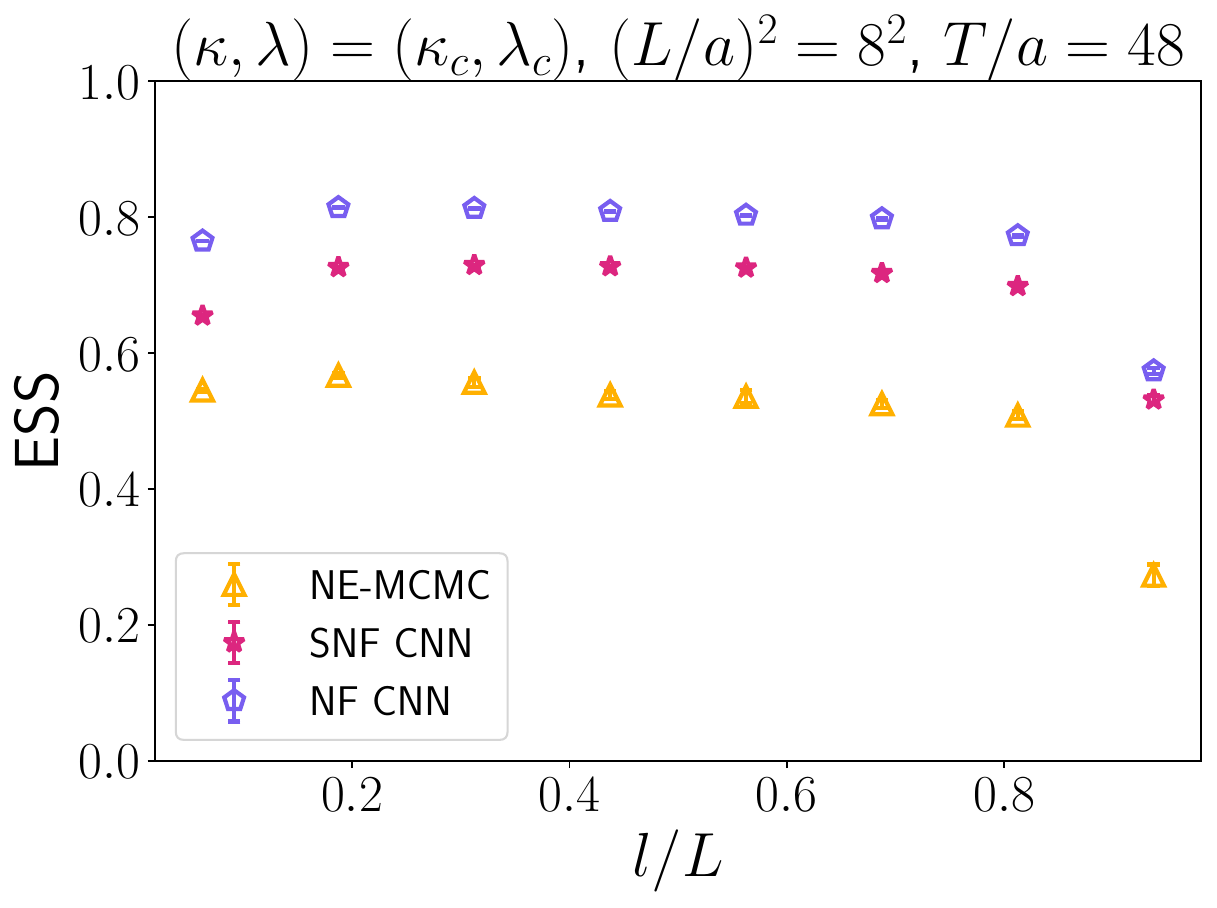}}
\caption{Models trained for $l/a=1$ and transferred to other values of $l/a$ in $D=1+1$ (left panel) and $D=2+1$ (right panel). The ESS is approximately constant for all the values of the length of the cut.} 
\label{fig:transfer_defect}
\end{figure*}

\textit{Design and training of the flows}: In what follows, we refer to a coupling layer, as it has been defined in the main text, as a transformation of the active part of the defect, leaving the frozen d.o.f. fixed; a block is a sequence of coupling layers after which every d.o.f. around the defect has been transformed once, in all the replicas. Specifically, our defect blocks are composed of a coupling layer per replica, therefore $2$ layers in the numerical experiments to compute $C_2$. In the $D=1+1$ case, we fixed $n_{step}=2$ for the NE-MCMC and the SNFs, with the latter architectures including $2$ blocks as well; for the NFs we used $4$ blocks. In the coupling layer of eq.~\eqref{eq:real_NVP} we used both fully-connected (FCNN) and convolutional (CNN) neural networks; the size of the patch on which the transformation acts is $4\times 5$ in all the cases except for the CNN-based NF, for which we set it to be $2\times 3$, as the one depicted in Fig.~\ref{fig:lattice_replica_space}. 
In $D=2+1$, we fixed $n_{step}=5$, while $5$ blocks are present both in the SNF and the NF. Here, we tested only CNNs to be able to transfer the model to larger volumes, as we are acting on a stripe of size $2\times 3\times L/a$.
The FCNN are shallow nets, without hidden layers, while CNNs, of size $3\times 3$ in $D=1+1$, $3\times3\times3$ in $D=2+1$, have three hidden layers with eight kernels for each of them. In order to enforce $\mathbb{Z}_2$ equivariance, we used a Tanh as non-linear activation and no bias. In the last layer of each neural network the weights are initialized to zero, such that the untrained coupling layer corresponds to the identity. In $D=1+1$, we trained the models on a lattice of two replicas with size $T\times L/a^2 = 128\times 16$, $l/a=1$ and $\kappa = \kappa_c = 0.2758297$, $\lambda=\lambda_c=0.03$~\cite{Bosetti:2015lsa}; the NFs and the FCNN-based SNF have been trained for $10^5$ steps, and the CNN-based SNF for $4\times 10^5$ steps, see Fig.~\ref{fig:comparison_training}. In $D=2+1$ all models have been trained for $2\times 10^5$ steps on a two-replicas lattice of size $T\times L^2/a^3=32\times 8^2$, $l/a=1$ and $\kappa=\kappa_c=0.18670475$, $\lambda=\lambda_c=0.1$~\cite{Hasenbusch:1999mw}. Notice also that the structure of an SNF allows for a particular kind of training~\cite{Matthews:2022sds}, where each block is trained separately. We found this strategy more convenient to train the $(2+1)$-dimensional SNFs, while all the others have been trained in one shot. The optimizer we used is ADAM~\cite{Kingma:2014vow}. For a summary of the design and training of the different models see Table~\ref{tab:flows_design} and Table~\ref{tab:flows_training}. 
All the trainings and samplings have been done on CINECA Leonardo accelerated clusters, using a single NVIDIA Ampere A100 GPU.

\textit{Data analysis}: For the estimation of the error we used the gamma method as implemented in pyerrors~\cite{Joswig:2022qfe}.

\textit{Comparison with the standard architecture}: In Fig.~\ref{fig:comparison_training} we show a comparison of the ESS of different models during training. In the horizontal axis, an era corresponds to 100 gradient updates and the ESS is averaged over the 100 steps. The models proposed in this work are compared with a standard RealNVP~\cite{Dinh:2017} architecture acting on the whole lattice of size $T\times L/a^2 = 128\times 16$ and two replicas, trained for $l/a = 1$ and at the critical point. Specifically, this last flow is made of four blocks of four coupling layer each, with both a masking on the replicas and an even-odd masking on the sites. The CNNs are made of three hidden layers with eight kernels each. After $6\times 10^5$ steps, the ESS of the standard architecture does not display any improvement.

\section{Study of the performances for different values of the parameters}

For an efficient study of the functional form of the entropic c-function it is crucial that the model, trained for a given size of the subsystem $A$, can be transferred to other values of $l/a$. At the same time, the operation of increasing the length of the cut by one lattice spacing, naming flowing from $l/a$ to $l/a+1$, is independent on length of the cut; indeed, as shown in Fig.~\ref{fig:transfer_defect}, both in $D=1+1$ and $D=2+1$, the models trained for $l/a$ can be transferred without retraining to essentially all the other length while retaining the same ESS, up to boundary effects.

\begin{figure*}[t]
\centerline{\includegraphics[height=0.25\textheight]{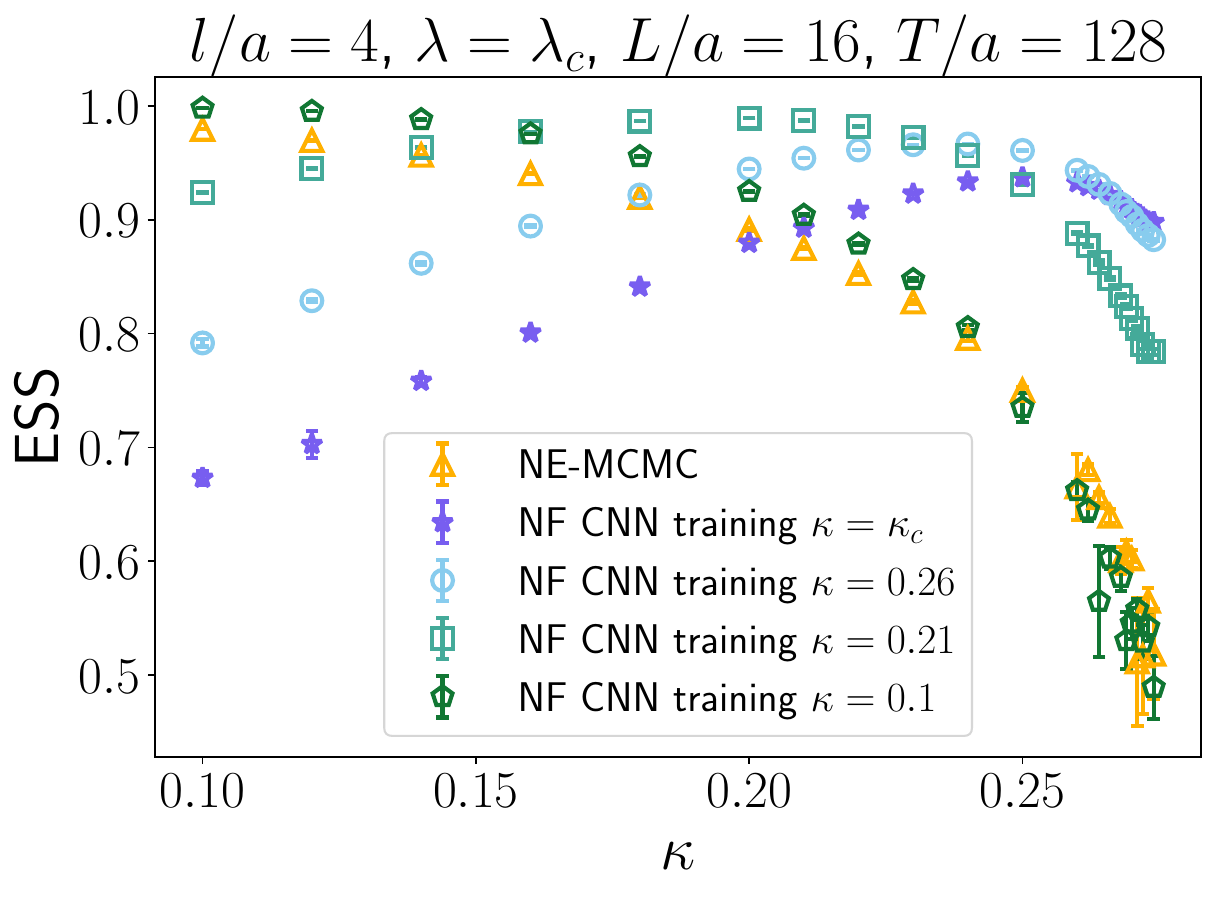}  \includegraphics[height=0.25\textheight]{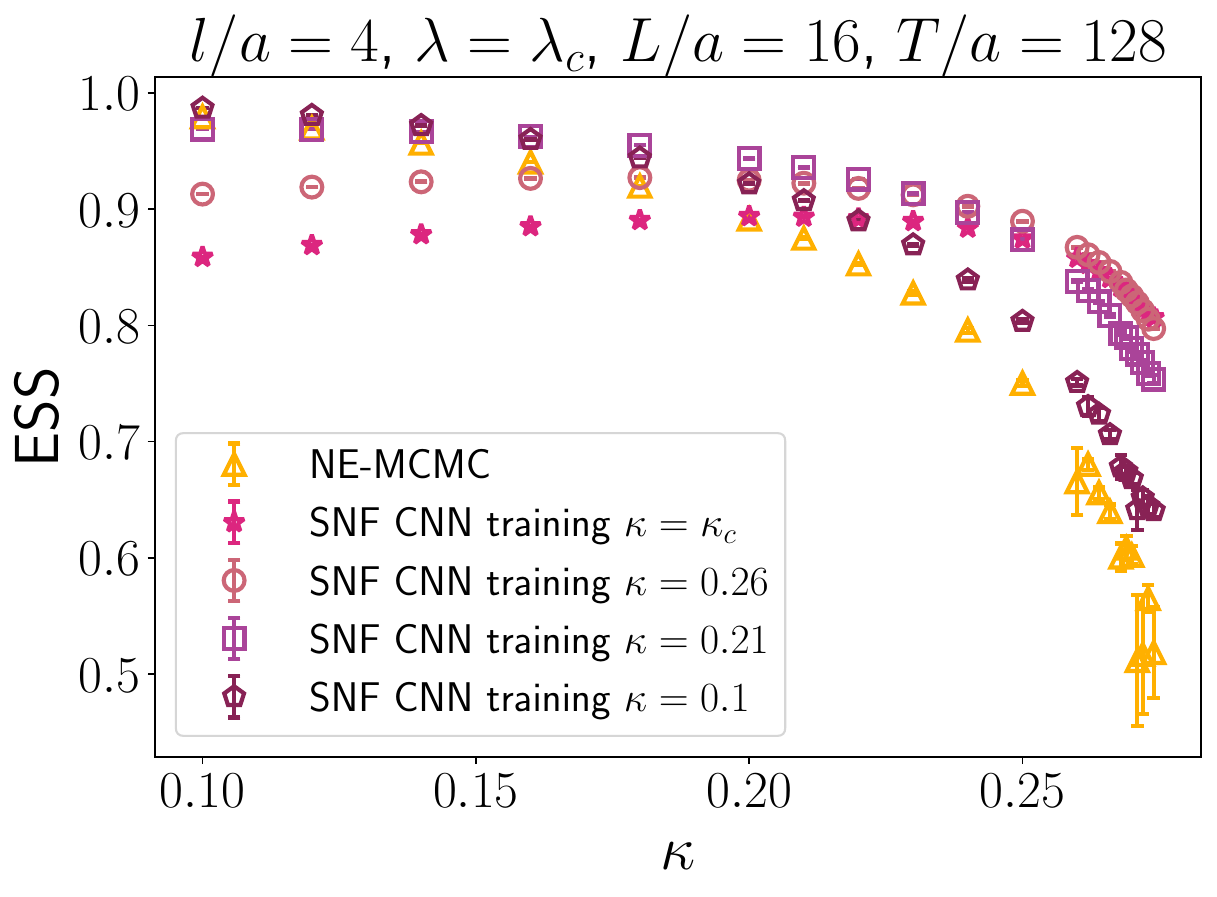}}
\caption{NFs (left panel) and SNFs (right panel) trained for different values of $\kappa$ and transferred without further retraining.} 
\label{fig:noncritical_kappa}
\end{figure*}

\begin{figure*}[t]
\centerline{\includegraphics[height=0.25\textheight]{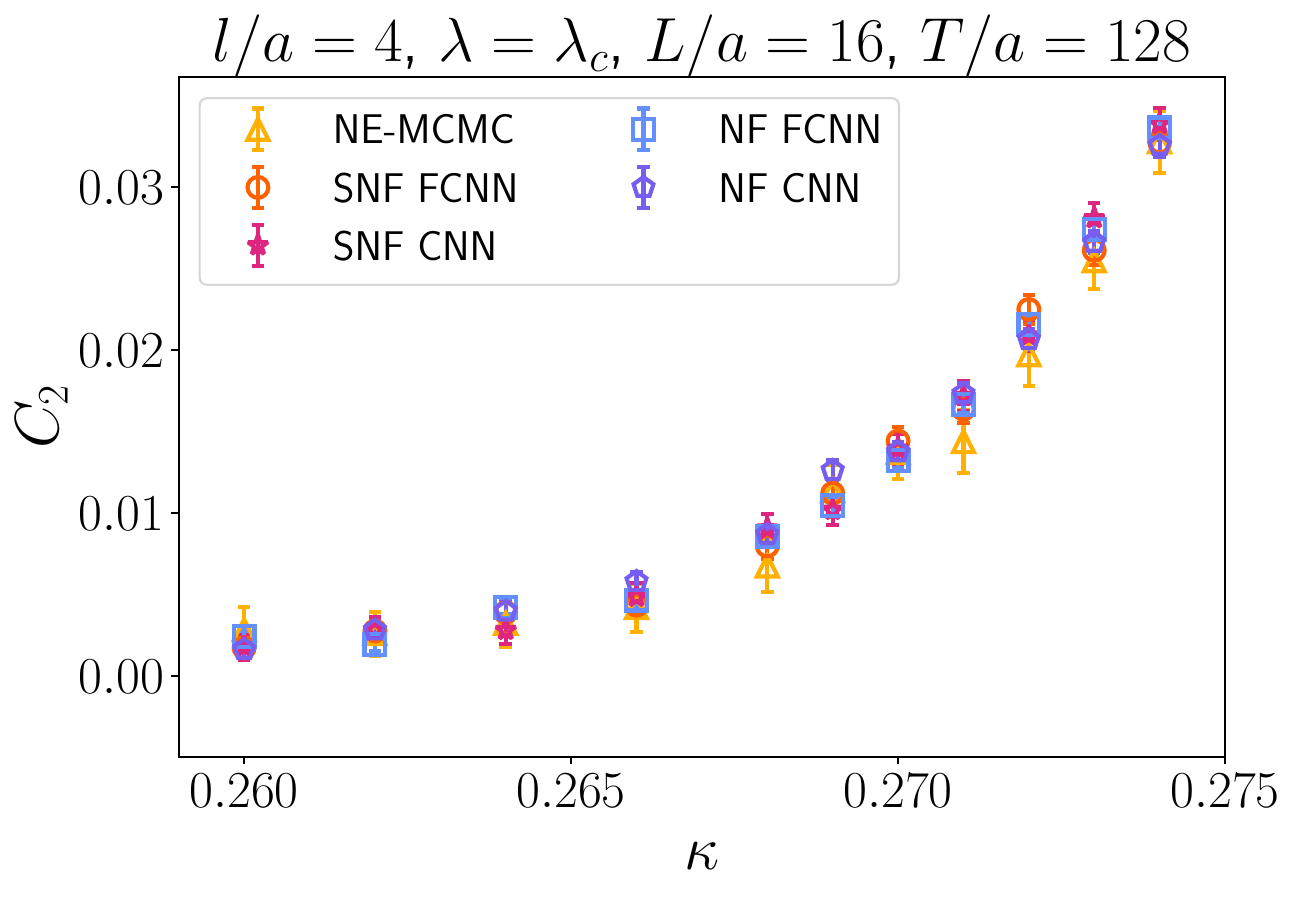}  \includegraphics[height=0.25\textheight]{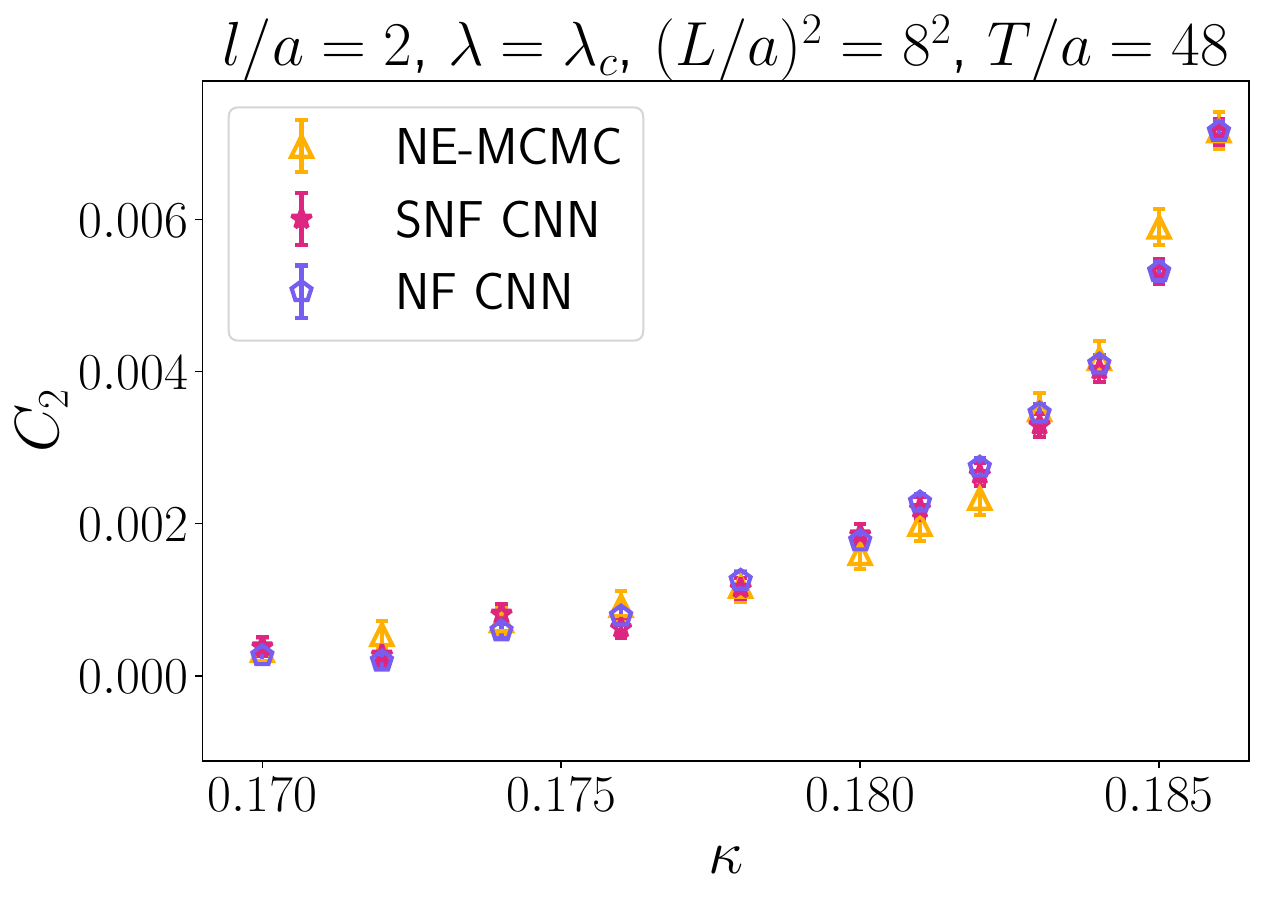}}
\caption{$C_2$ as a function of $\kappa$ in $D=1+1$ (left panel) and $D=2+1$ (right panel). Each model is trained at $\kappa=\kappa_c$ and sampling is performed at any $\kappa$ on the horizontal axis.} 
\label{fig:cfunction_for_different_kappa}
\end{figure*}

While the models we studied in the main text have been trained for a critical point of the $\phi^4$ theory, it is also clear that the range in $\kappa$ where they can be used for a reliable sampling extends well beyond the critical point. An interesting question is then if it is possible to perform the training in a different region of the phase diagram of the theory, away from the criticality, and then transfer \textit{across criticality} while retaining similar performances. In situations where critical slowing down is particularly severe, this would allow for a faster training on uncorrelated, yet non-critical, configurations. The picture emerging from Fig.~\ref{fig:noncritical_kappa} is particularly promising: both NFs and SNFs model trained for $\kappa=0.26$ and $\kappa=0.21$ have a large ESS when transferred to the critical point. In particular, SNFs are more robust under changes in the hopping, likely due to the underlying stochasticity---coming from the MC steps---, while NFs have in general a larger ESS close to the point where they have been trained. The possibility of training the model that far from $\kappa_c$ is particularly remarkable. For instance, for $\kappa=0.21$ the entropic c-function is practically zero (see Fig.~\ref{fig:cfunction_for_different_kappa}) and the physics is (almost) trivial compared to the critical point. Crucially, one can thus train the generative model in trivial regimes and use them to study the physics at criticality (where the physics gets more relevant).  

\begin{figure*}[t]
\centerline{\includegraphics[height=0.25\textheight]{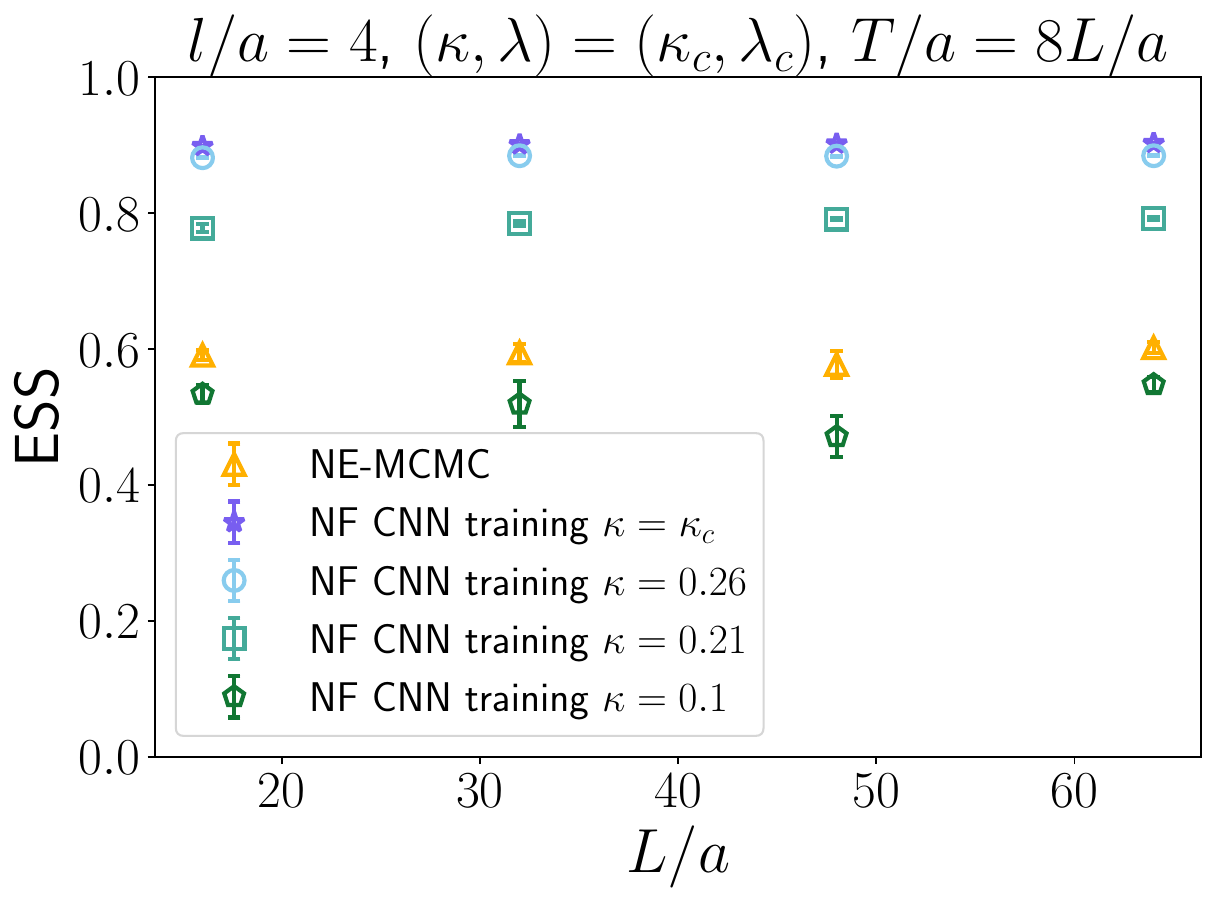}  \includegraphics[height=0.25\textheight]{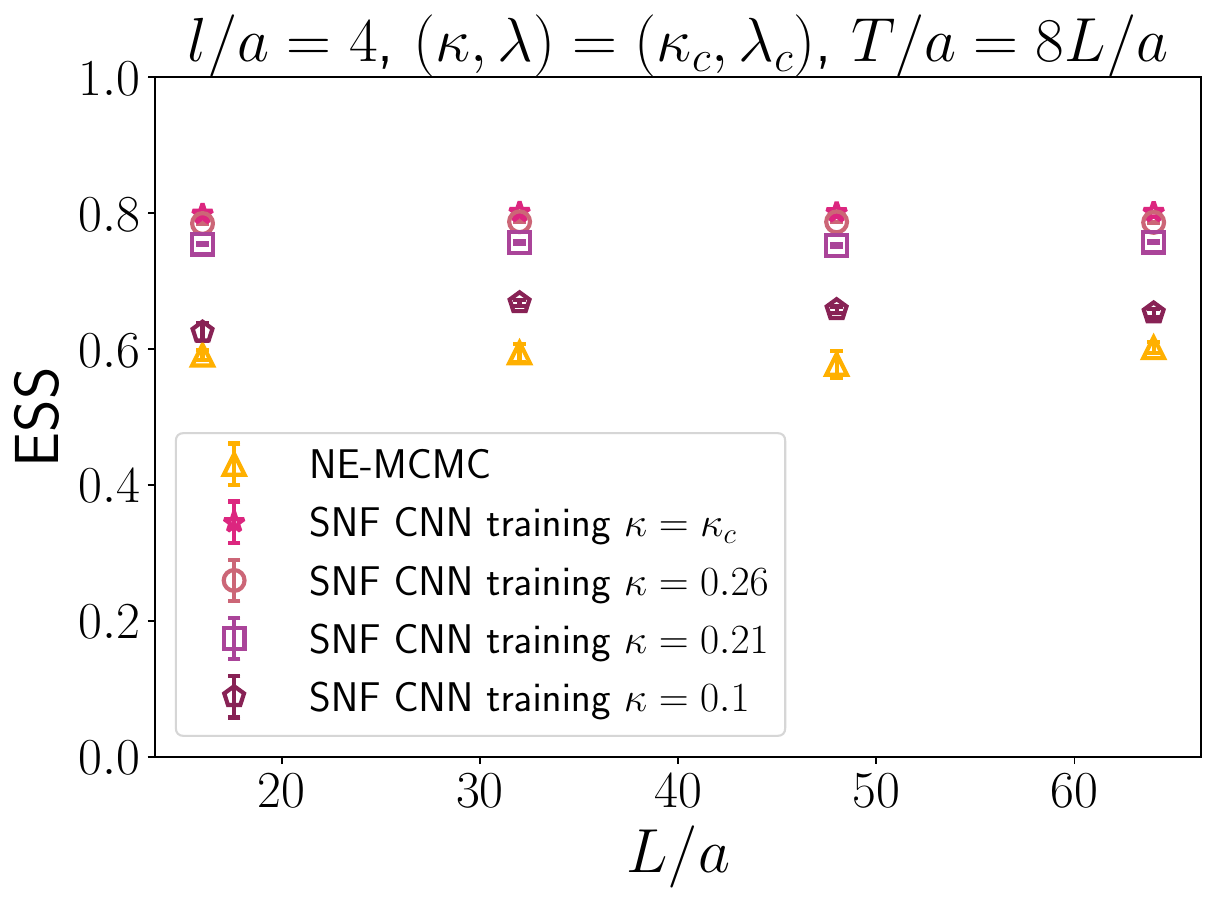}}
\caption{Transfer in the volume at $\kappa = \kappa_c$ of NFs (left panel) and SNFs (right panel) trained for different values of $\kappa$. Results in $D=1+1$} 
\label{fig:noncritical_volume}
\end{figure*}

The same model trained \textit{away from the criticality} can then be transferred in the volume, while setting $\kappa=\kappa_c$, as depicted in Fig,~\ref{fig:noncritical_volume}. Even though the correlation length of the system is progressively growing, the ESS remains constant.

The possibility of transferring the models in the phase space of the $\phi^4$ theory makes the flow-based samplers a useful tool to study the off-critical behavior of the entropic c-function. In particular, as shown in Fig.~\ref{fig:cfunction_for_different_kappa}, both in $D=1+1$ and $D=2+1$, $C_2$ is significantly different from zero in an interval close to the criticality, where NFs outperform all the other methods. In the present work we only studied the symmetric phase, to prevent any problem arising from mode collapse~\cite{PhysRevD.108.114501,Schuh:2025fky}. We leave to future studies the interesting and relevant question if and to what extent defect coupling layers suffer from mode collapse.

\section{Scaling}
\label{app:scaling}

\begin{figure*}[t]
\centerline{\includegraphics[height=0.22\textheight]{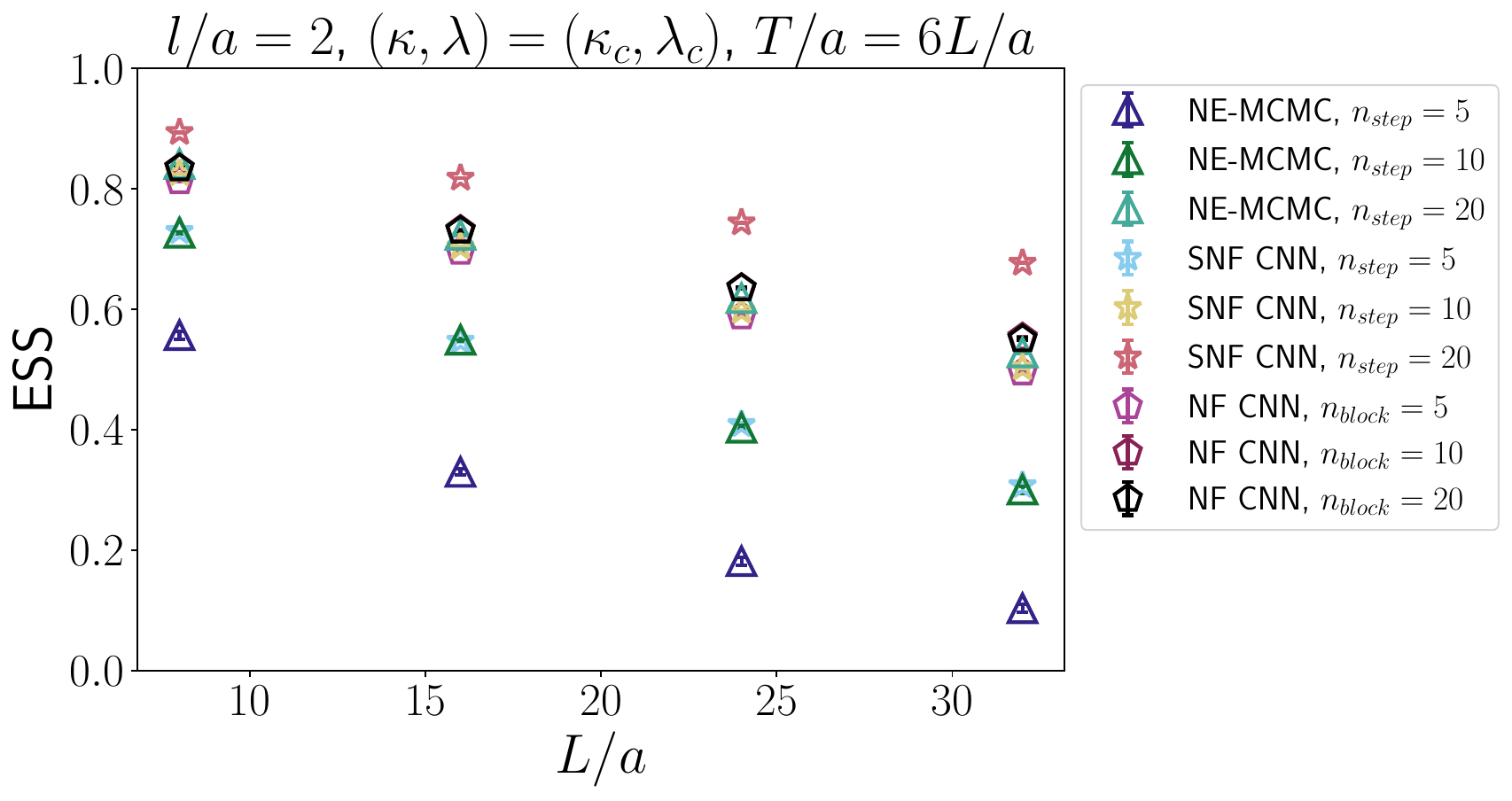}  \includegraphics[height=0.22\textheight]{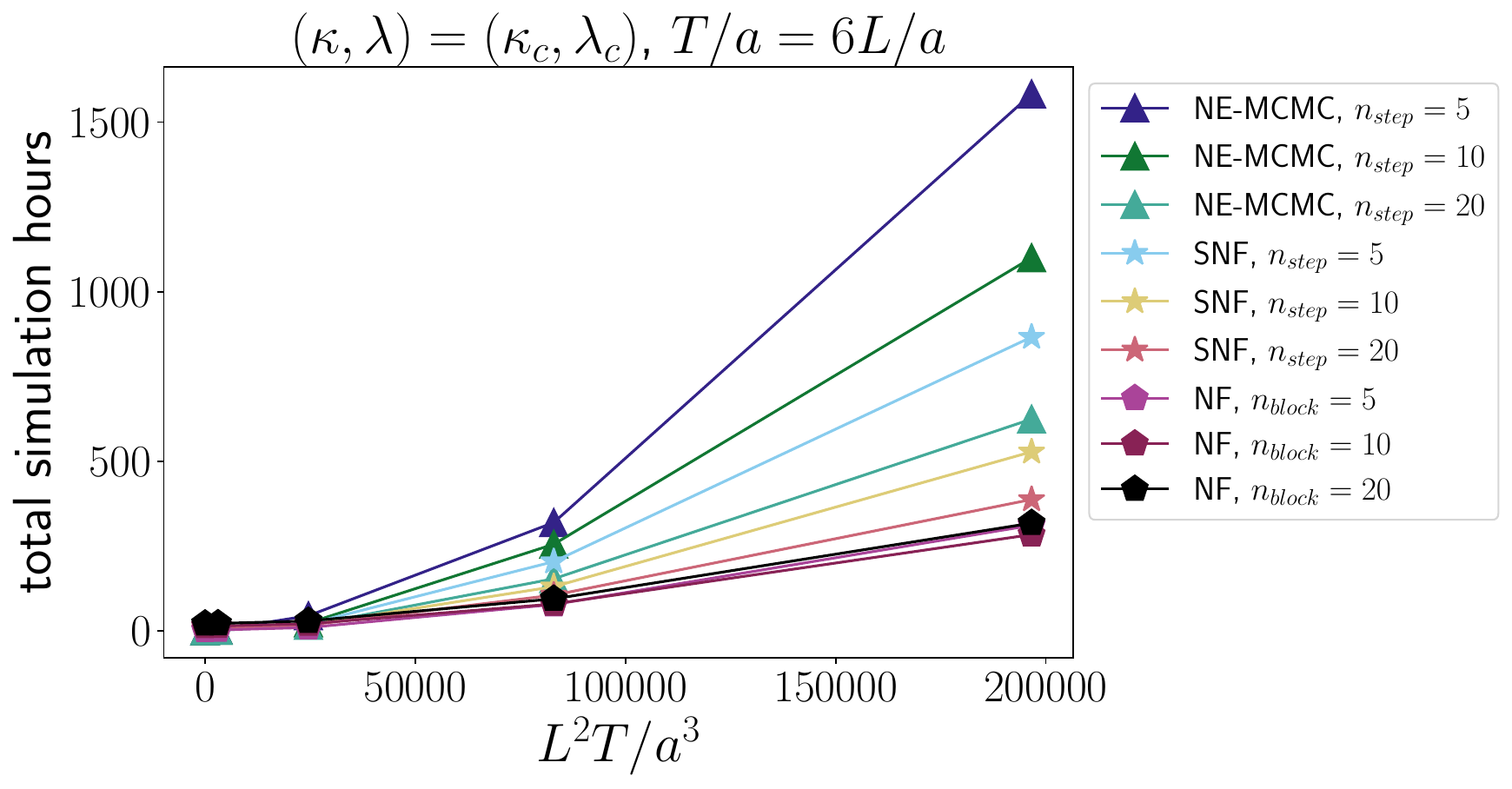}}
\caption{Comparison of the different algorithms in $D=2+1$ as the number of stochastic and deterministic (coupling) layers is increased. Left panel: transfer in volume. Right panel: total cost of the simulations.} 
\label{fig:scaling_nstep}
\end{figure*}

For the volumes we studied in the present work, the NFs discussed in the main test outperform the stochastic-based approaches. At the same time, both NE-MCMC and SNFs offer better scaling with the number of d.o.f.~\cite{Bonanno:2024udh,Bulgarelli:2024brv,Bulgarelli:2024cqc}. Even though a full study of the scaling is beyond the purpose of this work, here we present a preliminary analysis of different flow-based samplers, varying the number of stochastic and deterministic layers. In particular, in the left panel of Fig.~\ref{fig:scaling_nstep} we compare the same models we studied in the main text in $D=2+1$ with other, deeper, flows. Specifically, we study three NE-MCMC and three SNFs with $n_{step}=5,10,20$, and the same number of stochastic steps and deterministic blocks for the SNFs, and three NFs with $n_{block} = 5,10,20$. The picture emerging for the NE-MCMC and SNFs is similar: as the flows becomes deeper, the ESS increases, consistently with the expected scaling of both algorithms. Notice that all the three SNFs have been trained for the same number of gradient descent steps, in particular $2\times 10^5$. For NFs, instead, a larger number of blocks does not necessarily implies a larger ESS; in this case, we trained the NF with $n_{block}=5$ for $2\times 10^5$ steps, while the deeper flows made of $10$ and $20$ blocks have been trained for $8\times 10^5$ steps. Even for longer training times and a larger number of layers, the ESS results, for all three models, are close to each other, and do not display a clear scaling. The situation is similar if one analyzes the different methods in terms of the total cost (Fig.~\ref{fig:scaling_nstep}, right panel). More precisely, what is depicted here, as well as in Fig.~\ref{fig:cost_3d}, is the cumulative cost to compute the entropic c-function at the critical point of the $\phi^4$ theory in $D=2+1$; with cumulative, we mean that, for a given value of the volume, we plot the cost of the simulation for that specific lattice size as well as the ones for smaller lattices considered. The cost includes training, thermalization and sampling. As discussed in the main text, also here the NFs outperform all the other methods; the reason is due to the combination of two different factors: on one hand, the ESS of the NFs is larger than most of the other stochastic-based approaches, therefore it is possible to reach a target accuracy with less samples. On the other hand, NFs are made of only deterministic blocks, which are much faster than the stochastic layers. This explains why, even though the SNF with $n_{step}=20$ has the larger ESS (see left panel), in terms of a cost-performance comparison is still outperformed by all the other NFs. It is worth noticing, however, a hierarchy in the cost curves of the stochastic-based approaches: increasing the length of the flow leads to better performances. The same does not happen for the NFs. In Fig.~\ref{fig:scaling_nstep} (right) it becomes evident how the cost curves of the three NFs roughly collapse, showing a lack of a clear scaling.

\end{document}